\documentclass{pas}

\usepackage{multirow}
\usepackage{graphicx}	
\usepackage{amsmath}	
\usepackage{dsfont}
\newcommand{\mathbfss}[1]{\boldsymbol{#1}}

\begin{document}

\lefttitle{Publications of the Astronomical Society of Australia}
\righttitle{Wilensky \& Chokshi}

\jnlPage{1}{4}
\jnlDoiYr{2021}
\doival{10.1017/pasa.xxxx.xx}

\articletitt{Research Paper}

\title{Inferring radio beams from sparse sampling patterns: a case study with the Murchison Widefield Array}

\author{Michael J. Wilensky$^{1,\dag,\ast}$ \& Aman Chokshi$^{1,2,\ddag,\ast}$}

\affil{$^1$Department of Physics and Trottier Space Institute, McGill University, 3600 University Street, Montreal, QC H3A 2T8, Canada, $^2$The University of Melbourne, School of Physics, Parkville, VIC 3010, Australia, $^{\dag}$CITA National Fellow, $^{\ddag}$TSI Postdoctoral Fellow, $^{\ast}$Both authors contributed equally.}

\corresp{Aman Chokshi, Email: aman.chokshi@unimelb.edu.au}

\citeauth{Wilensky M J and Chokshi A, Inferring radio beams from sparse sampling patterns: a case study with the Murchison Widefield Array. {\it Publications of the Astronomical Society of Australia} {\bf 00}, 1--12. https://doi.org/10.1017/pasa.xxxx.xx}

\history{(Received xx xx xxxx; revised xx xx xxxx; accepted xx xx xxxx)}

\begin{abstract}
Accurate primary-beam models are essential for high-dynamic-range radio experiments, particularly searches for the redshifted 21-cm signal from the Epoch of Reionisation, where beam errors can couple bright foreground emission into the faint cosmological measurement. Using reprocessed satellite observations we infer the 137-MHz, zenith-pointed power beams of Murchison Widefield Array receiving elements. We develop two complementary Bayesian approaches for quantifying departures from nominal electromagnetic simulations: a physically motivated model of complex variations among individual dipoles and a hierarchical spline model for flexible direction-dependent structure. The measurements reveal shallower and displaced nulls together with coherent sidelobe deformations. The dipole model captures much of this broad structure, whereas the spline recovers additional localised features while remaining largely robust to isolated pass-shaped artefacts. Despite their sparse and distinct sampling patterns, four independent 20-day subsets, each containing approximately 20\% of the full dataset, recover consistent null and sidelobe morphology across the visible hemisphere. These results demonstrate that opportunistic satellite transmissions can support practical, repeatable beam reconstruction, while motivating future chromatic, phase-referenced measurements of the full Jones response.
\end{abstract}


\maketitle

\section{Introduction}

Scientific analysis of radio astronomy measurements depends on understanding the electromagnetic properties of the receiving elements. Typically these are formalized in terms of Jones matrices \citep{Jones1941}, which describe the (polarized) far-field radiation pattern of the receiving antenna. By time reversal symmetry of Maxwell's equations, this also describes the reception pattern i.e. how excited the receiving element is by a source, and is therefore an essential ingredient for inverting measurements into images of the radio sky. This pattern is often referred to as the \textit{beam} of the antenna. Models of the beam are used for various purposes, such as setting the overall flux scale of a radio map \citep{Jonas1998, Remazeilles2015} and in calibration procedures based on forward modeling the apparent sky signal. High dynamic range experiments that depend on forward modeling like cosmological 21cm global signal extraction and power spectrum estimation will fail if beam models are not sufficiently accurate \citep{EwallWice2017, Spinelli2022, Chokshi2024, Cumner2024, Sims2025}, and beam modeling is therefore of paramount importance for such experiments. 

A beam is often discussed in terms of three morphological regions. The bulk of the effective solid angle of the beam is usually contained in a \textit{main lobe}. Separating the main lobe and the \textit{sidelobes} are areas of low sensitivity referred to as \textit{nulls}, and there may be multiple groups of sidelobes separated by various nulls. While the size and shape of the main lobe can often be roughly predicted based on geometric considerations of the element (e.g. the shape of a dish reflector and the degree of underillumination), the sidelobe and null structures can depend sensitively on factors such as ground spill and coupling from nearby antennas in an interferometric array \citep{Fagnoni2021b, Gueuning2026}. It is typical to model the beam using a simulation based on a Maxwell solver \citep[e.g.][]{Fagnoni2021a, Cumner2022}. These solvers can produce highly detailed beam models, but their computational expense means they generally cannot account for the actual practical circumstances of a radio astronomy experiment such as mechanical idiosyncrasies between different antennas in an interferometer \citep{Chokshi2021} or the exact soil conditions at site \citep{Spinelli2022}, which may vary seasonally. This suggests an alternate approach to beam modeling, where the beam is measured or otherwise inferred from data taken with the instrument.  

Most beam measurements involve the use of a single extremely bright source at a time with variable position to probe the beam pattern. For example, transiting astrophysical sources draw circular tracks through the beam, and a source transiting boresight can be sufficient for a coarse measurement of the primary beam shape based on a single slice \citep{Haslam1974}. An accompanying antenna capable of tracking can be used to back out the beam along transits drawn by sources at multiple elevations using a technique called holography \citep{Iheanetu2019, Asad2021, CHIME2022}. The Sun can also be used to get data at multiple elevations due to its seasonal wandering \citep{Amiri2022}. For an array, one can infer beam patterns from visibility measurements with a sufficiently accurate sky model, sometimes called direction-dependent calibration \citep{Yatawatta2021, CHIME2022, Brackenhoff2025, Kern2025, Wilensky2025}. Non-astrophysical calibrator sources include transmitters attached to drones \citep{echo, Herman2024, Kuhn2025, Tyndall2025}, as well as satellite-based transmitters. 

Artificial satellites provide a complementary class of beam calibrator compared to celestial sources. Communication, weather, and navigation satellites can act as bright, narrow-band transmitters in the far field of a radio telescope~\citep{Neben_ORBCOMM_2015,Neben_HERA_2016, Line2018, Chokshi2021, Berger_GNSS_CHORD_2026}. Unlike celestial sources, whose apparent tracks are fixed by their declinations, satellites can cross the beam over a wide range of azimuths and elevations. Their trajectories are predictable from their orbital parameters, and repeated passes can collectively sample much of the visible hemisphere. The resulting coverage is in many cases irregular and incomplete, motivating inference methods that reconstruct a continuous beam while propagating uncertainty from the unsampled regions. The rapid growth of satellite constellations and CubeSat missions may provide further opportunities for beam measurements across multiple frequencies, including through unintended out-of-band emissions, while also creating a growing need to characterise and mitigate their impact on radio-astronomical observations \citep{Grigg_SatDetection_2025, Grigg_SatRFI_2025}.

The Murchison Widefield Array \citep[MWA;][]{Tingay_MWA_2013,
Wayth_MWA_2018, Tingay_MWA_2026} is a precursor to the Square
Kilometre Array \citep[SKA;][]{Mellema_SKA_2013}. Located at the Inyarrimanha Ilgari Bundara, the CSIRO Murchison Radio-astronomy Observatory in Western Australia, the MWA comprises 256 receiving tiles operating over the 80--300 MHz frequency range. Each tile contains 16 dual-polarised bow-tie dipoles arranged in a regular $4\times4$ grid above a $5\times5$ m conducting ground screen. Electronic beam steering is achieved using an analogue delay-line beamformer, which applies relative delays to the dipole signals before coherently combining them \citep{Neben_beamformer_2016}. Accurate modelling of the resulting tile primary beam is essential for calibration, imaging, and precision measurements of the low-frequency radio sky.

In this paper, we use reprocessed satellite-based measurements from \citet{Chokshi2021} to demonstrate Bayesian beam modeling methods, including inferring beams from sparse sampling patterns. These measurements were taken of ``power beam'' patterns (power absorbed from an unpolarized source) for 14 MWA receiving tiles. We build and compare two statistical models, a physically motivated one based on the MWA Average Embedded Element pattern \citep[AEE;][]{Sutinjo_FEE_2015}, and a spline-based hierarchical model. Bayesian statistical analysis \citep{Jaynes, Gelman2025} of these models allows us to quantitatively propagate uncertainty from noise and incomplete sampling to the inferred model parameters. This in turn allows us to make definitive statements about the nature of the data such as which features in the measurements are likely to be artefacts, and which subset of features are distinctly and reproducibly captured by each model. We focus the main analysis on one representative tile--polarisation combination, and repeat major components of the analysis for a second tile and polarisation, in Appendix \ref{sec:ant_10_demo}

The scientific contributions of this work are as follows. First, we demonstrate statistical constraints complete with propagated uncertainty on a highly interpretable physical beam model for the MWA at a single frequency. By itself, this can be used to speculate about plausible causes for differences compared to idealized electromagnetic simulations. If similar measurements can be done at more frequencies, the methodology we provide can be directly used to update the full Jones pattern as used in scientific processing pipelines with the MWA. Next, we identify deformations of the beam pattern that are not captured by the physical model, but are consistently captured by the spline model over different subsets of the data and therefore unlikely to be measurement artefacts. Since the spline model relates to a power beam, not a Jones matrix element, quantitative use of this model is limited to circumstances where the polarized response of the instrument is negligible (and again, where sufficient frequencies have been measured). However, the inferences still serve to demonstrate meaningful differences in the MWA beam from simulations, and are therefore important to understand more completely in future work to characterize the MWA beam for HDR experiments such as the search for the cosmic 21cm reionization signal \citep{Beardsley2016, Barry2019, Li2019, Trott2020, Rahimi2021, Wilensky2023, Nunhokee2025}. Finally, we demonstrate that these methods can be used to infer beams, including statistical uncertainties, from sparse sampling patterns in general. This is therefore a transferable methodology that extends beyond this particular use-case with satellite-based maps of the MWA beam into e.g. drone-based maps of other instruments' beams, though models will need to be at least somewhat tailored to the instrument at hand. 

The rest of the paper is laid out as follows. In \S\ref{sec:data}, we discuss how we have updated the pre-processing relative to \citet{Chokshi2021}, including recalibration and noise variance estimation. We formally describe the statistical models in \S\ref{sec:models}, and demonstrate their use with a reference MWA element in \S\ref{sec:results}. We discuss prospects for future experiments and analyses in \S\ref{sec:discussion}, and we state our main conclusions in \S\ref{sec:conclusion}. Finally, we offer several appendices with supplementary information including demonstration with another MWA tile and polarization (Appendix \ref{sec:ant_10_demo}), and about model selection and performance (Appendices \ref{sec:BIC} and \ref{sec:hierarchical_structure}).

\section{Data}
\label{sec:data}

The beam measurements used in this work were obtained by \citealt{Chokshi2021}, who measured the dual-polarised beam response of 14 tiles in the compact MWA core using transmissions from communication and weather satellites active in the 137-138~MHz band. Observations were conducted between September 2019 and March 2020, with the majority of continuous data  acquired in the first three months. A set of co-located reference antennas, consisting of a single MWA dipole mounted on an identical ground screen, were observed simultaneously to remove variations in satellite transmission power and the common propagation path. Each satellite pass therefore provided a calibrated cross-sectional slice through the tile beam, and the accumulation of thousands of passes produced all-sky beam maps for both instrumental polarisations. We describe our data preprocessing in the following subsections, pointing out differences from \citet{Chokshi2021} where relevant. In this work, we focus only on the zenith-pointed MWA beam. We expect some techniques we employ to extend to other pointings as-is, while others will require modification.

\subsection{Time-domain Reprocessing}

Raw satellite power measurements were recorded using RF Explorer spectrum analysers\footnote{\url{https://rfexplorer.com}}. Variations in the hardware produced native sampling rates of approximately 6 or 9~Hz. \citet{Chokshi2021} interpolated these measurements onto a common 1-Hz time grid and applied two successive second-order Savitzky--Golay filters~\citep{savgol_2002}, with window lengths of 11 and 15 samples. This processing introduced correlations between adjacent measurements, which would require a covariance model to propagate the measurement uncertainty formally through the beam inference. We are explicitly interested in the statistical uncertainty of the beam inference, so these correlations cannot simply be neglected.

Rather than model correlations introduced by avoidable interpolation and smoothing, we returned to the raw measurements and downsampled them onto a common 5-Hz time grid by selecting the nearest recorded measurement at each time step. No interpolation or additional filtering was applied. The resulting time series retains five times as many measurements as the previous 1-Hz product, providing additional leverage to characterise measurement noise and time-dependent systematics while avoiding the correlations introduced by the previous processing.

\subsection{Recalibration}
\label{sec:calibration}

As discussed by \citet{Chokshi2021}, satellite trajectories through the primary beam exceeded the linear operating range of the RF Explorer receivers, causing compression of the recorded signal. At the highest received powers, the compression reached approximately 10~dB, substantially distorting the measured primary beam response. \citet{Chokshi2021} corrected this effect using a global quadratic calibration derived from all receivers, reducing the peak residuals to approximately 2–4~dB near beam centre. In the present work, we revisit the original measurements and derive independent compression calibrations for each tile and RF Explorer. This reduces the residual discrepancy in the compressed regime to between approximately -0.5~dB and +0.8~dB relative to the MWA fully embedded element beam model \citep[FEE;][]{Sokolowski_FEE_2017}, while preserving the low-power sidelobe measurements used throughout the remainder of this work.

\begin{figure}
    \centering
    \includegraphics[width=\linewidth]{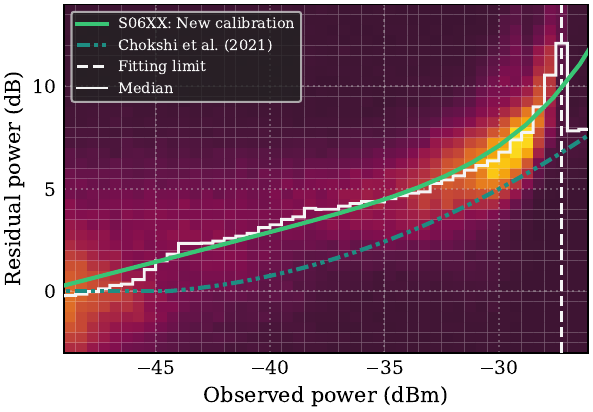}
    \caption{RF Explorer compression recalibration for tile S06XX. The two-dimensional histogram shows the density of primary-beam measurements as a function of recorded power and the residual relative to the FEE prediction. The stepped white curve gives the median residual in $0.5~\mathrm{dBm}$ bins. The solid green curve shows the receiver-specific monotonic convex calibration fitted to these medians, while the dashed teal curve shows the global quadratic calibration of \citet{Chokshi2021} derived using all 14 measured tiles. Measurements to the right of the vertical dashed line belong to a sparsely populated high-power sequence and are excluded from the fit. The increasing residual towards high recorded powers reflects compression of the RF Explorer response.}
    \label{fig:recalibration}
\end{figure}

The recalibration is derived using satellites passes that transit the primary beam, where the FEE model has been shown to accurately reproduce the beam response \citep[][]{Sokolowski_FEE_2017}. For each tile and RF Explorer, we compare the measured beam response to the corresponding FEE prediction as a function of the recorded RF Explorer power. Figure \ref{fig:recalibration} shows representative two dimensional histograms of the residual power relative to the FEE model (dB) as a function of absolute received power (dBm). The median residual demonstrates what appears to be a convex monotonic relationship characteristic of receiver compression.

This procedure only assumes that the FEE model provides an adequate description of the primary beam in the high-power regime used for calibration. True discrepancies between the beam and the FEE model are expected to primarily be a function of observing direction, whereas receiver compression depends only on the recorded power. Consequently, physical beam deviations contribute scatter about the median compression relation rather than biasing the derived calibration curve. Since only measurements in the primary beam exceed the linear operating range of the RF Explorers, the recalibration modifies only these high-power samples, leaving the lower-power sidelobe measurement that form the basis of the present beam inference effectively unchanged.

To extract this convex monotonic relationship, we fit a constrained model of Bernstein polynomials to the data, restricting ourselves to data that are greater than -50 dBm in absolute power. We observed a secondary cluster of measurements at extremely high power values (typically higher than -25 dBm) that display a different monotonic relationship in this plane than the rest of the data. This secondary cluster is a minority of the data (a few percent). Rather than model it separately, we exclude it based on a density cut. Included data are to the left of the vertical dashed line in Figure \ref{fig:recalibration}. Rather than fit directly to the data, we fit to the median log-ratio as a function of log-power using bins of width 0.5 dBm. This helps avoid interference in the fit from the secondary cluster.

Bernstein polynomials have several properties that make them convenient for guaranteeing monotonicity and convexity. The $n + 1$ Bernstein polynomials of degree $n$,  defined on the real interval $[0,1]$, form a basis for polynomials of degree $n$. The $\nu$'th basis element is given by
\begin{equation}
    b_{\nu, n}(x) = \binom{n}{\nu}x^\nu(1-x)^{n-\nu}
\end{equation}
Since these basis elements are all positive, a superposition of Bernstein polynomials with positive (negative) coefficients is necessarily positive (negative). That is, if $a_\nu > 0$ for all $\nu$ in $\{0,1,..n\}$, then 
\begin{equation}
    \sum_{\nu=0}^na_\nu b_{\nu, n}(x) > 0
\end{equation}
for all $x\in[0,1]$. Furthermore, using the fact that derivatives of Bernstein polynomials are superpositions of Bernstein polynomials, one can show that if the $a_\nu$ are ordered, that is $a_0 \leq a_1 \leq ... \leq a_n$ then the polynomial will be monotonic \citep{GHOSAL2023}. Combining these facts, we can define a strictly convex and monotonic model function, $m(x)$, by demanding that 
\begin{equation}
    m(x) \equiv c + \int_0^x \mathrm{d}x' \sum_{\nu=0}^n a_\nu b_{\nu, n}(x')
    \label{eq:bernstein_model}
\end{equation}
where $a_\nu$ are positive and ordered and $c$ is a constant. This is to say, if the derivative of $m(x)$ is positive and increasing everywhere, then $m(x)$ is by definition monotonically increasing and convex. The advantage of using expression $\ref{eq:bernstein_model}$ is that the integrals can be done analytically by noting that the $\nu$'th basis element is the probability density function of a Beta random variable. This means the definite integral above is given by the corresponding cumulative density functions, which are commonly implemented function in scientific computation packages such as \texttt{Scipy} (which we used). 

To apply this to our data, we rescaled and shifted the absolute power values so that they were contained in the interval $[0,1]$. A (for example) naive least-squares fit of Bernstein polynomials to the resulting data would not guarantee positivity or orderliness of the coefficients. To guarantee these properties, we reparametrized the $a_\nu$ by defining positive $c_k$ such that
\begin{equation}
    a_\nu = \sum_{k=0}^\nu c_k
\end{equation}
i.e. the $a_\nu$ are the cumulative sum of the $c_k$. We then solved a bounded least-squares optimization in terms of $c_k$, with $n = 10$. For our use case, bounding the $c_k$ on the interval $[0, 20]$ was sufficient for effective fits, but other use cases may vary. We set the integration constant in Equation \ref{eq:bernstein_model} to 0, equivalent to assuming the amplifier is exactly linear at the lower edge of the fit interval. As illustrated in Figure \ref{fig:recalibration}, this is not precisely obeyed by all the data. However, in the vast majority of cases this was an excellent assumption. 

We also noticed that for some tiles, the very highest powers (say the highest 1 dBm bin) could not be precisely captured by the fit without tailoring various fit hyperparameters (such as the number of polynomial terms) to the tile at hand. Consequently, there are a minority of slightly miscalibrated passes in the data, which we ultimately believe to cause residual pass-shaped artefacts in the main lobe measurements. We explore this in \S\ref{sec:results}, but have generally found the recalibrated data to be vastly superior in the main lobe due to the per-tile calibration making up for the large systematic error from the global fitting procedure, illustrated by the discrepancy between solid and dash-dot lines in Figure \ref{fig:recalibration}. We developed calibration solutions for all 14 tiles analyzed by \citet{Chokshi2021}, but only fully examine two tiles in this work.

\subsection{Binning to HEALPix}

Theoretically the beam models described in \S\ref{sec:models} can be fit directly to the calibrated pass data. However, many passes cover nearly the same trajectories across the beam, meaning that their information content with respect to the model is highly redundant. With this dataset, computational expense can be reduced by multiple orders of magnitude for essentially the same model fit by first binning the passes to pixels on the hemisphere and keeping careful track of the per-pixel statistics. We use a HEALPix \citep{Healpix} binning with \texttt{NSIDE}~$= 32$, corresponding to approximately $1.8^\circ$ resolution.

\begin{figure}
    \centering
    \includegraphics[width=\linewidth]{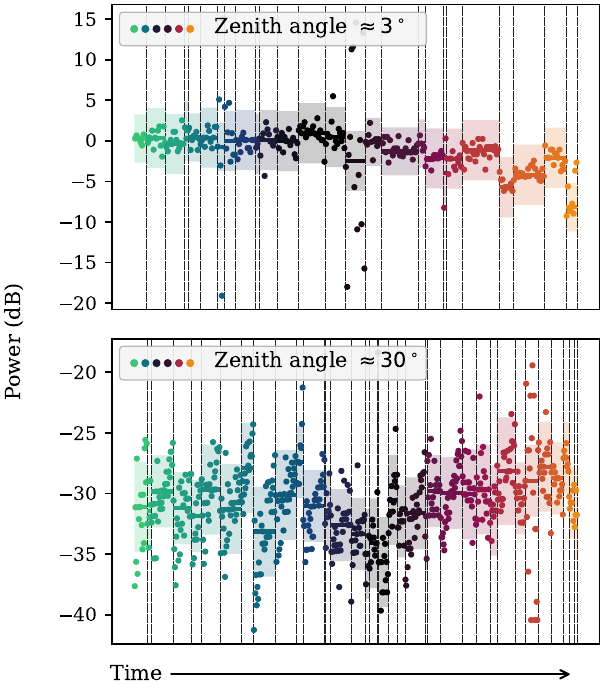}
    \caption{Variation among satellite passes contributing to two representative HEALPix pixels centred at zenith angles of approximately $3^\circ$ (top) and $30^\circ$ (bottom). Points show the individual $5$-Hz power measurements, with colours distinguishing successive passes and vertical grey lines marking their boundaries. The shaded region for each pass shows its median power together with the estimated within-pass noise. Although all measurements in each panel occupy the same HEALPix pixel, their central values vary between passes because of differences in their sub-pixel trajectories and any remaining pass-dependent systematics. The horizontal axis concatenates the measurements in time order and is discontinuous across pass boundaries.}
    \label{fig:passes_in_a_pixel}
\end{figure}

The primary beam exhibits small variations over the extent of a single pixel, and the exact geometry of a given pass through a pixel can change its median value (for example). Since we measure the beam in logarithmic units and there is sizable per-sample noise in the passes (typically a few dB), this usually presents as different passes within a pixel having different constant offsets with respect to one another. For pixels in primary beam nulls, there are noticeable gradients in the passes even in the span of one pixel (Figure \ref{fig:passes_in_a_pixel}). The variation in central value of passes within a pixel is an additional source of variation in the data on top of receiver noise that must be accounted for to accurately characterize the statistics of the fit residuals (what we call noise in \S\ref{sec:models}). We formally account for this as follows.

For pixels where gradients are negligible, the $j$th pass within a pixel can be modelled as 
\begin{equation}
    \mathbfss{d}_j = \mu_j \mathds{1}_{N_i} + \mathbfss{n}_j
\end{equation}
where $\mathbfss{d}_j$ is a vector holding the $N_j$ measured data points for this pass, $\mu_j$ is a per-pass constant, $\mathds{1}_{N_j}$ is a vector of ones with length equal to $N_j$, and $\mathbfss{n}_j$ is a vector of independent and identically distributed (i.i.d.) noise with variance $\sigma^2_j$. The exact distribution of $\mu_j$ depends on the distribution of pass geometries as well as the distribution of any unknown varying systematics that we may have failed to calibrate. Supposing for simplicity that the $\mu_j$ are i.i.d. per pass with variance across passes of $\sigma_\mu^2$, the covariance matrix of the $P$ passes, $\mathbfss{d} = (\mathbfss{d}_1^\top, \mathbfss{d}_2^\top, ...\mathbfss{d}_P^\top)^\top$, in hypothetical repeated experiments has a block-diagonal structure. The $j$th diagonal block is given by
\begin{equation}
    \mathbfss{C}_j \equiv \sigma_u^2 \mathds{1}_{N_j}\mathds{1}_{N_j}^\top + \sigma^2_j\mathbfss{I}_{N_j}
\end{equation}
where $\mathbfss{I}_{N_j}$ is the identity matrix of dimension $N_i$. The variance of a sample average for a pixel would then be
\begin{equation}
\begin{aligned}
    \sigma^2_\mathrm{pix} &= \frac{1}{(\sum_j N_j)^2} \sum_{k=1}^P \sigma_\mu^2 (\mathds{1}_{N_k}^\top \mathds{1}_{N_k})^2 + \sigma_k^2 \mathds{1}_{N_k}^\top \mathds{1}_{N_k} \\
    &= \frac{1}{(\sum_j N_j)^2} \sum_{k = 1}^P N_k^2 \sigma_\mu^2 + N_k \sigma_k^2.
\end{aligned}
    \label{eq:noise_full_gore}
\end{equation}

As a sanity check, consider the case where $N_j = K$ and $\sigma_j^2 = \sigma^2$. Then $N \equiv \sum_j N_j = PK$, giving
\begin{equation}
\begin{aligned}
    \sigma^2_\mathrm{pix} &= \frac{PK^2}{(PK)^2}\sigma_\mu^2 + \frac{PK}{(PK)^2}\sigma^2 \\
    &= \frac{\sigma_\mu^2}{P} + \frac{\sigma^2}{N}
\end{aligned}
\end{equation}
which is the result we would expect with i.i.d. noise per-sample and an i.i.d. offset per-pass. Our noise estimation follows the more complete Equation \ref{eq:noise_full_gore} since we have significant variation between number of points-per-pass in a given pixel.

\begin{figure}
    \centering
    \includegraphics[width=\linewidth]{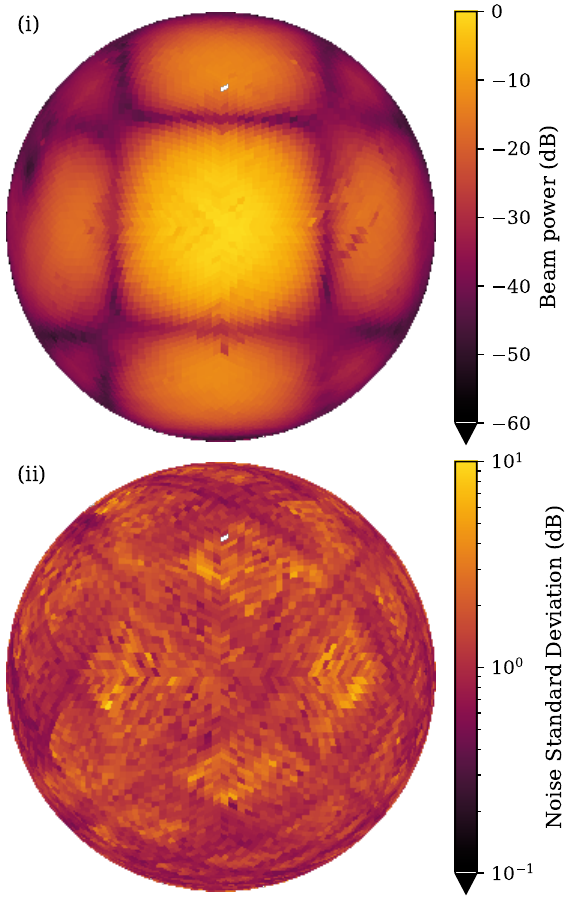}
    \caption{HEALPix-binned beam measurements and corresponding noise estimates for the full dataset from the East-West oriented dipole in the MWA Southern compact hex (S06XX), shown in orthographic projection with $\texttt{NSIDE}=32$. Panel~(i) shows the median beam power in each pixel. Panel~(ii) shows the per-pixel noise standard deviation calculated using Equation~\eqref{eq:noise_full_gore}, displayed on a logarithmic colour scale. The noise is heteroscedastic because the number of measurements, within-pass noise, and variation between contributing passes differ across the hemisphere. White pixels were not sampled by the measured satellite passes. }
    \label{fig:noise_map}
\end{figure}

Some passes appear to have heavier-tailed noise than other passes. Therefore, we use median-based estimators for all quantities in Equation \ref{eq:noise_full_gore}. Specifically, we get the per-pass variance in a pixel using the median absolute deviation. While binning, we collect the median of each pass, and use the median absolute deviation of the per-pass medians to estimate $\sigma_\mu$. This gives us a noise map to use when fitting models. An example noise map is shown in Figure \ref{fig:noise_map}. The final value in each pixel is given by the median of all contributions to that pixel. Note we have ignored the effects of gradients, mainly observable in beam nulls. We expect this to have a negligible effect on the binned data and noise maps since medians are not strongly sensitive to gradients. A more precise treatment might, for example, fit all passes in a given pixel to a small inclined plane, and take the central value of the plane.

Before inference, we apply two additional pixel-level quality cuts. First, pixels containing fewer than three contributing measurements are excluded because their within-pixel noise statistics cannot be estimated reliably. Second, we guard against spuriously small noise estimates in pixels sampled by only one satellite pass. This becomes relevant for the 20-day subsets considered in \S\ref{sec:time_jackknife}, where a small number of single-pass pixels were assigned implausibly small uncertainties, occasionally reaching $\sigma_{\mathrm{pix}}\sim10^{-5}$--$10^{-4}~\mathrm{dB}$. They can arise when the within-pass MAD is anomalously small, for example when a pixel contains a short or nearly constant sequence of samples. These values are not interpreted as exceptionally precise measurements, but as systematically erroneous passes that must be excluded. Such pixels exert disproportionate influence on the likelihood, causing numerical difficulties for the Bayesian sampling methods described in \S\ref{sec:models} and likely causing large systematic fit errors even in cases where the sampling methods converge. 

For the single-pass pixels, we calculate a robust $z$-score from $\log_{10}\sigma_{\mathrm{pix}}$, using the median and gaussian-scaled median absolute deviation across the single-pass pixels as the location and scatter estimators, respectively. We exclude pixels lying more than three standard deviations below the median. Restricting this cut to single-pass pixels preserves genuinely precise estimates supported by multiple independent passes. Neither quality cut removes any pixels from the full dataset. In the 20-day subsets, the low-uncertainty cut removes fewer than $1\%$ of the retained pixels in every case.

\section{Models for Inference}
\label{sec:models}

We analyze these data using two different models. One model is a parametrized extension of the Average Embedded Element model \citep[AEE;][]{Sutinjo_FEE_2015}. The AEE model is based on electromagnetic simulations of the MWA receiving element, and our parametrized extension is constrained to have certain physical features. While a physically motivated model has the major advantage of interpretability, it is unable to capture all the structure of the measured beams. We compare this model to a second model based on splines. This model is regularized in a way that allows us to capture smooth structure in the data unreachable by the physical model but without overfitting the data. Its structure is of general use and may extend to other instruments with some tailoring of certain model choices. In the following subsections, we first explain previous relevant MWA beam models, and then the models that we use for statistical analysis in this work.

\subsection{MWA Beam Models}

The MWA beam model has undergone several generations of development: the Analytic, Average Embedded Element (AEE), and Full Embedded Element (FEE) models. These models differ primarily in their descriptions of the individual dipole responses and their treatment of electromagnetic interactions within the tile \citep{Sutinjo_FEE_2015,Sokolowski_FEE_2017}. The Analytic and AEE models share a factorised description of the tile response, whereas the FEE model relaxes the assumption that every dipole has the same element pattern. We introduce this common formalism before describing the successive models in detail.

In the factorised description, the tile Jones response is written as
\begin{equation}
    \mathbf{J}_{\mathrm{tile}}(\hat{\mathbf{s}},\nu)
    =
    \mathbf{A}_{xy}(\hat{\mathbf{s}},\nu)\,
    g(\hat{\mathbf{s}},\nu)\,
    \mathbf{J}_e(\hat{\mathbf{s}},\nu).
    \label{eq:aee_tile_beam}
\end{equation}
where $\mathbf{J}_e$ is the direction- and frequency-dependent Jones response of an individual element, $g$ describes the ground-plane response, and
\begin{equation}
    \mathbf{A}_{xy}(\hat{\mathbf{s}},\nu)
    =
    \begin{pmatrix}
        f_x(\hat{\mathbf{s}},\nu) & 0\\
        0 & f_y(\hat{\mathbf{s}},\nu)
    \end{pmatrix}
    \label{eq:array_factor_matrix}
\end{equation}
contains the array factors of the two tile polarisations \citep{Sutinjo_FEE_2015}. This separation distinguishes the response common to each element from the interference produced by combining signals across the tile.

For a zenith-pointed tile, the array factor for polarisation
$p\in\{X,Y\}$ is
\begin{equation}
    f_p(\hat{\mathbf{s}},\nu)
    =
    \frac{1}{16}
    \sum_{n=0}^{15}
    i_{n,p}(\nu)\,
    e^{i\mathbf{k}\cdot\mathbf{r}_n},
    \label{eq:current_array_factor}
\end{equation}
where $i_{n,p}$ is the complex current at port $p$ of the $n$th dual-polarised dipole, $\mathbf{r}_n$ is its position relative to the tile centre, and $\mathbf{k}=(2\pi\nu/c)\hat{\mathbf{s}}$. Each array factor therefore depends both on the geometrical arrangement of the dipoles and on their complex port currents.

The 32 port currents are collected into the vector $\mathbf{i}$ and are related to the applied excitation voltages $\mathbf{v}$ by
\begin{equation}
    \mathbf{i}
    =
    \left(\mathbf{Z}+\mathbf{Z}_L\right)^{-1}\mathbf{v},
    \label{eq:aee_port_currents}
\end{equation}
where $\mathbf{Z}$ is the $32\times32$ mutual-impedance matrix of the tile and $\mathbf{Z}_L$ is the diagonal matrix of LNA input impedances. The off-diagonal entries of $\mathbf{Z}$ describe coupling both between dipoles and between their orthogonal ports. This coupling allows the port currents to differ between dipoles even when the applied excitation voltages are uniform.

The Analytic MWA beam model approximates each bow-tie antenna as an idealised short dipole above a conducting ground plane and neglects mutual coupling. In this limit, the mutual-impedance matrix is diagonal and uniform excitation produces identical port currents. The current-weighted array factors in Equation~\eqref{eq:current_array_factor} therefore reduce, up to an overall normalisation, to geometrical sums over the dipole positions.

The AEE model retains the factorised form of Equation~\eqref{eq:aee_tile_beam}, but replaces the idealised dipole response with an average embedded-element pattern obtained from electromagnetic simulations of the complete tile. It also incorporates mutual coupling through the impedance matrix in Equation~\eqref{eq:aee_port_currents}, giving each array factor its corresponding set of complex port currents. A single average element response is nevertheless used for every dipole.

The FEE model removes this common-element approximation. Each port is assigned its own simulated embedded response, and the tile beam is constructed by coherently combining the responses of all 32 ports. The FEE response therefore cannot, in general, be separated into a common element pattern and the current-weighted array-factor matrix of Equation~\eqref{eq:aee_tile_beam}. This more complete treatment substantially improves the modelling of the MWA's direction-dependent polarimetric response, particularly at higher frequencies where mutual-coupling effects become more pronounced \citep{Sutinjo_FEE_2015,Sokolowski_FEE_2017}.

\begin{figure}
    \centering
    \includegraphics[width=\columnwidth]{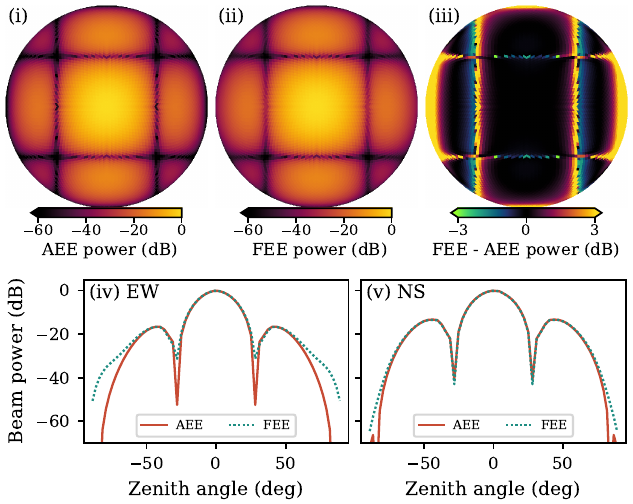}
    \caption{Comparison of the zenith-pointed MWA AEE and FEE beam models at $137~\mathrm{MHz}$ for the XX polarisation. Panels~(i) and (ii) show the peak-normalised AEE and FEE power beams, respectively, in orthographic projection. Panel~(iii) shows the difference between the FEE and AEE beams in dB. The difference scale is limited to $\pm3~\mathrm{dB}$ to reveal structure across the main beam; values outside this range are saturated. Panels~(iv) and (v) show east--west and north--south slices through zenith. Negative zenith angles denote west and south, respectively. The models agree closely across the main lobe but differ increasingly around the nulls and towards the horizon.}
    \label{fig:aee_fee_comparison}
\end{figure}

Figure~\ref{fig:aee_fee_comparison} compares the zenith-pointed AEE and FEE power beams at $137~\mathrm{MHz}$. The models agree closely across the high-response region but diverge progressively towards the nulls and horizon. The east--west and north--south slices make these differences particularly clear, showing that relatively small differences across the main lobe develop into appreciable differences in the null depths and low-response beam profiles. Where both beams exceed $-20~\mathrm{dB}$, their log-ratio has a median below $0.01~\mathrm{dB}$ and an RMS of $0.232~\mathrm{dB}$. The RMS log-ratio increases to $0.614$, $1.334$, and $2.346~\mathrm{dB}$ when the comparison is extended to responses above $-30$, $-40$, and $-50~\mathrm{dB}$, respectively. The two models therefore describe the main beam similarly, but make appreciably different predictions in the lower-response regions.

The two statistical models developed below use these electromagnetic models for different purposes. We construct the dipole model from the AEE because its factorised form provides a simple separation between the common embedded-element response and the current-weighted array factor. This permits the effects of dipole-level complex gains to be introduced directly and interpreted in terms of changes to the nominal port currents. An analogous parametrisation could potentially be developed from the port-dependent responses of the FEE, but would be more complex and less readily interpretable. By contrast, the spline model does not require a dipole-level factorisation and instead parametrises flexible deviations from a fixed reference beam. We therefore use the FEE as its reference because it provides the more complete nominal electromagnetic description.

The two approaches consequently address complementary questions. The dipole model asks how much of the measured beam variation can be represented by changes to the dipole gains within the AEE formalism, whereas the spline model identifies additional spatial structure supported by the measurements relative to the FEE prediction. Both ultimately predict the measured power beam, but their inferred deviations are defined relative to different nominal models.

\subsection{Dipole Model}

The dipole model parametrises tile-to-tile beam variations through complex gains associated with the individual dipoles. The two orthogonal tile polarisations are modelled independently. For each polarisation, the model begins from the AEE description and applies a complex gain to the directly driven contribution to each dipole's port current, while the cross-coupled contribution from the orthogonal polarisation and the average embedded-element response remain fixed at their nominal values. This provides a compact, physically motivated model for dipole-level departures from the nominal tile beam. We first define the complex-gain parametrisation and its associated statistical model, and then examine the symmetries and identifiability of the resulting inference problem.

\subsubsection{Complex-Gain Parametrisation}

We construct the dipole model for a single tile polarisation $p\in\{X,Y\}$, with the orthogonal polarisation denoted by $\bar p$. The X and Y polarisations are fitted independently throughout this work.

The linear relation in Equation~\eqref{eq:aee_port_currents} allows the current at each port to be separated into contributions from the two applied excitations. For the $p$ port of dipole $n$, the AEE current is
\begin{equation}
    i_{n,p}^{\mathrm{AEE}}
    =
    I^{\mathrm{dir}}_{n,p}
    +
    I^{\mathrm{cross}}_{n,p},
    \label{eq:aee_current_decomposition}
\end{equation}
where $I^{\mathrm{dir}}_{n,p}$ is the current produced by directly driving polarisation $p$, and $I^{\mathrm{cross}}_{n,p}$ is the current induced at the same port by driving the orthogonal polarisation $\bar p$. The latter arises from electromagnetic coupling between the two sets of ports.

We associate each dipole with a complex gain
\begin{equation}
    \gamma_{n,p}
    =
    a_{n,p}e^{i\phi_{n,p}},
    \label{eq:complex_dipole_gain}
\end{equation}
where $a_{n,p}$ and $\phi_{n,p}$ describe the relative amplitude and phase of the directly driven current. The modified port current is then
\begin{equation}
    i_{n,p}(\gamma_{n,p})
    =
    \gamma_{n,p}I^{\mathrm{dir}}_{n,p}
    +
    I^{\mathrm{cross}}_{n,p}.
    \label{eq:modified_port_current}
\end{equation}
Thus, the gain modifies only the directly driven contribution. The cross-coupled contribution is retained at its nominal AEE value.

Substituting Equation~\eqref{eq:modified_port_current} into Equation~\eqref{eq:current_array_factor} gives the modified array factor
\begin{equation}
    f_p(\hat{\mathbf{s}},\nu;\boldsymbol{\gamma}_p)
    =
    \frac{1}{16}
    \sum_{n=0}^{15}
    \left[
        \gamma_{n,p}I^{\mathrm{dir}}_{n,p}
        +
        I^{\mathrm{cross}}_{n,p}
    \right]
    e^{i\mathbf{k}\cdot\mathbf{r}_n},
    \label{eq:dipole_model_array_factor}
\end{equation}
where $\boldsymbol{\gamma}_p=(\gamma_{0,p},\ldots,\gamma_{15,p})$. The array factor of the orthogonal polarisation remains fixed at its nominal AEE value. The resulting tile response is
\begin{equation}
    \mathbf{J}_{\mathrm{dip}}
    (\hat{\mathbf{s}},\nu;\boldsymbol{\gamma}_p)
    =
    \mathbf{A}_{xy}
    (\hat{\mathbf{s}},\nu;\boldsymbol{\gamma}_p)\,
    g(\hat{\mathbf{s}},\nu)\,
    \mathbf{J}_e(\hat{\mathbf{s}},\nu),
    \label{eq:dipole_model_jones}
\end{equation}
where only the diagonal entry of $\mathbf{A}_{xy}$ corresponding to the fitted polarisation is modified.

For later convenience, we define the effective complex coefficient contributed by dipole $n$ as
\begin{equation}
    c_{n,p}(\gamma_{n,p})
    =
    \gamma_{n,p}I^{\mathrm{dir}}_{n,p}
    +
    I^{\mathrm{cross}}_{n,p}.
    \label{eq:effective_dipole_coefficient}
\end{equation}
The fitted array factor can then be written compactly as
\begin{equation}
    f_p(\hat{\mathbf{s}},\nu;\boldsymbol{\gamma}_p)
    =
    \frac{1}{16}
    \sum_{n=0}^{15}
    c_{n,p}(\gamma_{n,p})
    e^{i\mathbf{k}\cdot\mathbf{r}_n}.
    \label{eq:effective_coefficient_array_factor}
\end{equation}

Equations~\eqref{eq:effective_dipole_coefficient} and \eqref{eq:effective_coefficient_array_factor} define the forward model relating the complex dipole gains to the tile beam. We next specify the likelihood and prior distributions used to infer these gains from the beam measurements.

\subsubsection{Statistical Model}

For a given tile and polarisation, the data consist of beam-power measurements $d_q$ and corresponding noise standard deviations $\sigma_q$ in HEALPix pixels indexed by $q$. Both are expressed in dB, with $\sigma_q$ derived using Equation~\eqref{eq:noise_full_gore} and illustrated in Figure~\ref{fig:noise_map}. Because the polarisations are modelled independently, we suppress their index on these quantities.

For polarisation $p$, the unnormalised model power response is
\begin{equation}
    P_p(\hat{\mathbf{s}};\boldsymbol{\gamma}_p)
    =
    \left[
        \mathbf{J}_{\mathrm{dip}}
        (\hat{\mathbf{s}};\boldsymbol{\gamma}_p)
        \mathbf{J}_{\mathrm{dip}}^\dagger
        (\hat{\mathbf{s}};\boldsymbol{\gamma}_p)
    \right]_{pp}.
    \label{eq:dipole_model_power}
\end{equation}
The beam model is normalised by its peak response, which occurs at zenith for the zenith-pointed beam considered here. The normalised power beam is therefore
\begin{equation}
    B_p(\hat{\mathbf{s}};\boldsymbol{\gamma}_p)
    =
    \frac{
        P_p(\hat{\mathbf{s}};\boldsymbol{\gamma}_p)
    }{
        P_p(\hat{\mathbf{s}}_{\mathrm{zen}};
        \boldsymbol{\gamma}_p)
    }.
    \label{eq:normalised_dipole_beam}
\end{equation}
The model prediction in pixel $q$, expressed in dB, is
\begin{equation}
    m_{q,p}(\boldsymbol{\gamma}_p)
    =
    10\log_{10}
    B_p(\hat{\mathbf{s}}_q;\boldsymbol{\gamma}_p).
    \label{eq:dipole_model_db}
\end{equation}
This normalisation sets the peak model response to $0~\mathrm{dB}$.

We model the data in each pixel as being corrupted by additive noise, independent per-pixel. Preliminary analysis of the data suggested that relative to a best fit model, the noise is non-Gaussian, with exponential tails. We make extensive use of the scaled ``soft Laplace'' distribution in both statistical models, which has probability density function
\begin{equation}
    p(z | s) = h(z; s) \equiv \frac{1}{\pi s}\mathrm{sech}(z/s).
    \label{eq:soft_laplace}
\end{equation}
where $s$ is its scale parameter. The standard deviation of this distribution is $\pi s/2$. We therefore convert the noise standard deviation $\sigma_q$ derived for each pixel into the corresponding soft-Laplace scale
\begin{equation}
    s_q
    =
    \frac{2\sigma_q}{\pi}.
    \label{eq:soft_laplace_noise_scale}
\end{equation}
This ensures that the standard deviation of the likelihood in each pixel matches the noise map derived in Section~\ref{sec:data}. Assuming independent residuals $r_{q,p}=d_q-m_{q,p}(\boldsymbol{\gamma}_p)$, the log-likelihood is
\begin{equation}
    \log p(\mathbf{d}\mid\boldsymbol{\gamma}_p)
    =
    -\sum_q
    \left\{
        \log\left(2\sigma_q\right)
        +
        \log\cosh
        \left[
            \frac{\pi}{2}
            \frac{
                d_q-m_{q,p}(\boldsymbol{\gamma}_p)
            }{
                \sigma_q
            }
        \right]
    \right\}.
    \label{eq:dipole_log_likelihood}
\end{equation}

We assign independent priors to the gain amplitudes and phases defined in Equation~\eqref{eq:complex_dipole_gain}. To permit relative amplitude variations without introducing an additional common scale, we constrain the mean amplitude to unity by writing
\begin{equation}
    a_{n,p}=16w_{n,p},
    \qquad
    \sum_{n=0}^{15}w_{n,p}=1.
    \label{eq:amplitude_simplex}
\end{equation}
The amplitude fractions $\mathbf{w}_p=(w_{0,p},\ldots,w_{15,p})$ therefore lie on the 15-dimensional simplex. We assign them the symmetric Dirichlet prior
\begin{equation}
    \mathbf{w}_p
    \sim
    \operatorname{Dirichlet}
    \left(\alpha_a\mathbf{1}_{16}\right),
    \label{eq:amplitude_dirichlet_prior}
\end{equation}
where $\alpha_a$ controls the concentration about the nominal AEE amplitudes, $a_{n,p}=1$. We adopt $\alpha_a=10$, which concentrates the prior around equal dipole amplitudes while allowing appreciable relative variations.

We allow relative phase variations but assume no prior preference for positive or negative offsets. To respect the periodicity of phase, we assign the offsets a zero-centred von Mises distribution,
\begin{equation}
    p(\phi_{n,p}\mid\kappa_\phi)
    =
    \frac{
        \exp\!\left(\kappa_\phi\cos\phi_{n,p}\right)
    }{
        2\pi I_0(\kappa_\phi)
    },
    \qquad n=1,\ldots,15,
    \label{eq:phase_von_mises_prior}
\end{equation}
where $I_0$ is the modified Bessel function of the first kind and $\kappa_\phi$ controls the concentration about zero. The limiting case $\kappa_\phi=0$ is uniform on the circle. We choose $\kappa_\phi$ to give the von Mises prior a circular standard deviation of $10^\circ$.

Because the measurements contain only power information, the likelihood is invariant under a common phase rotation of all effective dipole coefficients i.e. their overall phase is unidentifiable. We fix this freedom by choosing dipole 0 as the phase reference and setting
\begin{equation}
    \phi_{0,p}=0.
    \label{eq:phase_reference}
\end{equation}
Assuming independent amplitude and phase priors, the posterior is
\begin{equation}
    p(\mathbf{w}_p,\boldsymbol{\phi}_p\mid\mathbf{d})
    \propto
    p(\mathbf{d}\mid\boldsymbol{\gamma}_p)\,
    p(\mathbf{w}_p)\,
    p(\boldsymbol{\phi}_p).
    \label{eq:dipole_posterior}
\end{equation}
The simplex and phase-reference constraints set the common amplitude and phase conventions of the parametrisation. A discrete ambiguity nevertheless remains because of the conjugate-rotation symmetry of the tile, which we examine in the following section.

\subsubsection{Symmetry and Identifiability}
\label{sssec:symmetry}

Despite fixing the common amplitude and phase conventions, the dipole model retains a discrete non-identifiability because the measurements constrain the power beam rather than the complex electric-field response.

From Equations~\eqref{eq:dipole_model_jones} and \eqref{eq:dipole_model_power}, the power response of polarisation $p$ can be written as
\begin{equation}
    P_p(\hat{\mathbf{s}};\boldsymbol{\gamma}_p)
    =
    \left|
        f_p(\hat{\mathbf{s}};\boldsymbol{\gamma}_p)
    \right|^2
    \left|g(\hat{\mathbf{s}})\right|^2
    \left[
        \mathbf{J}_e(\hat{\mathbf{s}})
        \mathbf{J}_e^\dagger(\hat{\mathbf{s}})
    \right]_{pp}.
    \label{eq:factorised_dipole_power}
\end{equation}
All dependence on the inferred gains therefore enters through $\lvert f_p\rvert^2$. Using the effective coefficients defined in Equation~\eqref{eq:effective_dipole_coefficient}, the array factor is
\begin{equation}
    f_p(\hat{\mathbf{s}})
    =
    \frac{1}{16}
    \sum_{n=0}^{15}
    c_{n,p}\,
    e^{i\mathbf{k}\cdot\mathbf{r}_n},
    \label{eq:symmetry_array_factor}
\end{equation}
where we suppress the dependence on frequency and
$\boldsymbol{\gamma}_p$ for clarity.

The MWA tile is invariant under a $180^\circ$ rotation about its centre. With our dipole numbering, this maps $n\leftrightarrow15-n$, with
\begin{equation}
    \mathbf{r}_{15-n}=-\mathbf{r}_n.
    \label{eq:dipole_rotation}
\end{equation}
Consider the transformed coefficients
\begin{equation}
    c'_{n,p}
    =
    \lambda e^{i\alpha}c^*_{15-n,p},
    \qquad \lambda>0,
    \label{eq:normalised_coefficient_symmetry}
\end{equation}
where $\lambda$ and $\alpha$ are a common amplitude scale and phase, respectively. Their array factor is
\begin{align}
    f'_p(\hat{\mathbf{s}})
    &=
    \frac{\lambda e^{i\alpha}}{16}
    \sum_{n=0}^{15}
    c^*_{15-n,p}
    e^{i\mathbf{k}\cdot\mathbf{r}_n} \\
    &=
    \frac{\lambda e^{i\alpha}}{16}
    \sum_{m=0}^{15}
    c^*_{m,p}
    e^{-i\mathbf{k}\cdot\mathbf{r}_m} \\
    &=
    \lambda e^{i\alpha}f_p^*(\hat{\mathbf{s}}),
    \label{eq:array_factor_symmetry}
\end{align}
where the second line follows by setting $m=15-n$ and using Equation~\eqref{eq:dipole_rotation}. Consequently,
\begin{equation}
    \left|f'_p(\hat{\mathbf{s}})\right|^2
    =
    \lambda^2
    \left|f_p(\hat{\mathbf{s}})\right|^2.
    \label{eq:array_factor_power_symmetry}
\end{equation}
The transformed coefficients therefore multiply the unnormalised power response by the direction-independent factor $\lambda^2$. This factor cancels under peak normalisation, so the transformed coefficients produce the same normalised power beam, the same prediction in dB, and hence the same likelihood. Thus, a $180^\circ$ rotation and complex conjugation of the effective coefficient pattern leave the inferred power beam unchanged; its common amplitude and phase are likewise removed by the beam normalisation and power measurement, respectively.

Importantly, this symmetry acts on the effective coefficients $c_{n,p}$ rather than directly on the inferred gains $\gamma_{n,p}$. From Equation~\eqref{eq:effective_dipole_coefficient}, the gain corresponding to the transformed coefficient is
\begin{equation}
    \gamma'_{n,p}(\lambda,\alpha)
    =
    \frac{
        \lambda e^{i\alpha}
        \left[
            \gamma_{15-n,p}I^{\mathrm{dir}}_{15-n,p}
            +
            I^{\mathrm{cross}}_{15-n,p}
        \right]^*
        -
        I^{\mathrm{cross}}_{n,p}
    }{
        I^{\mathrm{dir}}_{n,p}
    }.
    \label{eq:transformed_gain}
\end{equation}
The symmetry-related gains are therefore not obtained by simply rotating and conjugating the original gains: the nominal direct and cross-coupled currents must also be included.

The transformed gains do not automatically satisfy the amplitude and phase conventions imposed by the statistical model. We choose $\lambda$ and $\alpha$ such that
\begin{align}
    \arg\gamma'_{0,p}(\lambda,\alpha) &= 0,
    \label{eq:transformed_phase_constraint}\\
    \sum_{n=0}^{15}
    \left|\gamma'_{n,p}(\lambda,\alpha)\right| &= 16.
    \label{eq:transformed_amplitude_constraint}
\end{align}
The first condition restores dipole 0 as the phase reference, while the second enforces the mean-unity amplitude constraint. After solving these coupled nonlinear constraints, the symmetry-related parameters are
\begin{equation}
    a'_{n,p}
    =
    \left|\gamma'_{n,p}\right|,
    \qquad
    \phi'_{n,p}
    =
    \arg\gamma'_{n,p}.
    \label{eq:transformed_amplitude_phase}
\end{equation}

This construction maps a gain solution onto a second point in the constrained parameter space with an identical predicted power beam. The two solutions are related by complex conjugation and a $180^\circ$ rotation of the effective dipole coefficients, followed by the global scaling and phase rotation needed to restore the amplitude and phase conventions. We therefore expect the likelihood, and potentially the posterior, to contain two symmetry-related modes. We test this prediction in Section~\ref{sec:results}.

\subsubsection{Implementation and Posterior Sampling}

To enable efficient gradient-based inference, we developed \texttt{mwa\_jaxbeam}\footnote{\url{https://github.com/amanchokshi/mwa_jaxbeam}}, a targeted JAX implementation of the 137-MHz, zenith-pointed MWA AEE model. We validated its XX and YY power responses against the AEE implementation distributed with \texttt{pyuvdata}~\citep{Hazelton_pyuvdata_2017,Keating_pyuvdata_2025}. Across the visible hemisphere, the two implementations agree to an RMS of approximately $0.02~\mathrm{dB}$ where both beam responses exceed $-60~\mathrm{dB}$. Automatic differentiation and just-in-time compilation provide the likelihood gradients and rapid forward evaluations required for posterior sampling.

\begin{figure}
    \centering
    \includegraphics[width=\linewidth]{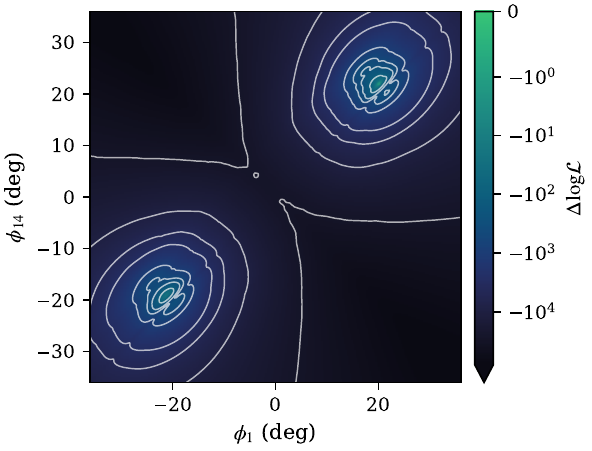}
    \caption{Likelihood slice through the complex dipole model for tile S06XX. The relative log-likelihood is evaluated as a function of $\phi_1$ and $\phi_{14}$, with all other phases fixed to zero and all amplitudes fixed to unity. The relative log-likelihood is measured from the maximum within the slice, and the contours span $\Delta\log\mathcal{L}=-30{,}000$ to $-100$. Two well-separated high-likelihood regions are visible, each containing narrower internal structure.}
    \label{fig:likelihood_slice}
\end{figure}

Initial inference tests indicated that the posterior could not be reliably explored using local gradient-based methods. Optimisation from different initial parameter values converged to distinct solutions, while independently initialised gradient-based Markov chain Monte Carlo (MCMC) chains remained confined to different regions of parameter space. To investigate the likelihood structure responsible for this behaviour, we evaluated two-dimensional slices through pairs of phase parameters. Figure~\ref{fig:likelihood_slice} shows the likelihood as a function of $\phi_1$ and $\phi_{14}$, with all dipole amplitudes fixed to unity and all other phases fixed to zero. It is therefore a slice through the likelihood rather than a marginalisation or optimisation over the remaining parameters.

The slice is strongly bimodal, consistent with the conjugate-rotation symmetry derived in the preceding section. Within each mode, the likelihood is concentrated along curved ridges and smaller isolated islands, separated by likelihood differences spanning many orders of magnitude. The resulting geometry is therefore substantially more complex than two smooth, approximately Gaussian modes.

We sample the posterior using annealing, which bridges an accessible initial distribution and a complex target through a sequence of intermediate distributions~\citep{Neal_Annealed_Sampling_2001}. Writing the amplitude and phase parameters collectively as $\boldsymbol{\theta}$, we use an inverse-temperature parameter $\beta_t$ to progressively introduce the likelihood, defining the intermediate target
\begin{equation}
    \pi_t(\boldsymbol{\theta})
    \propto
    p(\boldsymbol{\theta})
    p(\mathbf{d}\mid\boldsymbol{\theta})^{\beta_t},
    \qquad
    0=\beta_0<\cdots<\beta_T=1.
    \label{eq:annealed_smc_target}
\end{equation}
The sequence begins with the prior at $\beta_0=0$ and ends with the full posterior at $\beta_T=1$. We sample these intermediate targets using the sequential Monte Carlo (SMC) framework of \citet{DelMoral_SMC_2006}, implemented with \texttt{BlackJAX}\footnote{\url{https://blackjax-devs.github.io/blackjax}}. In this framework, an ensemble of parameter samples, termed particles, is evolved stochastically through the sequence of intermediate distributions until the final population represents samples from the posterior. The sampled parameter vector contains the 15 independent amplitude degrees of freedom and the 15 free relative phases, with the physical gains reconstructed so that the amplitude-simplex and phase-reference constraints are satisfied.

At low inverse temperatures, the broad prior dominates the tempered target and the large likelihood contrasts between modes are strongly suppressed. Regions that become effectively disconnected in the full posterior can therefore remain connected by regions of non-negligible probability, allowing the particle population to explore between emerging modes. As $\beta_t$ increases, the likelihood increasingly concentrates probability within these modes, and transitions between them become progressively less probable. This gradual evolution allows SMC to establish particle support across the modes before they separate and subsequently refine the distribution within each. Provided the particle population is sufficiently large, the final population can therefore represent posterior structure that would be difficult to recover using independently initialised local MCMC chains.

Each SMC stage begins with a population of particles representing $\pi_{t-1}$. For a candidate inverse temperature $\beta_t$, incremental reweighting assigns each particle an importance weight according to how its relative probability changes between successive targets. After these weights are normalised to sum to unity, the effective sample size is calculated as the inverse sum of their squares and decreases as the total weight becomes concentrated among fewer particles. \texttt{BlackJAX} selects $\beta_t$ adaptively so that the effective sample size is approximately $0.8N$, where $N$ is the number of particles. The resulting weighted population represents $\pi_t$, while the weights also estimate the ratio of normalising constants between the two successive targets. Accumulating these ratios over the complete annealing sequence yields the Bayesian evidence $Z=p(\mathbf{d})$. We then apply systematic resampling. The normalised cumulative weights divide the unit interval into segments whose lengths are proportional to the particle weights. We draw a random offset uniformly from $[0,1/N)$ and use it to position $N$ points separated by $1/N$ across the unit interval. Each grid point falls within one of these segments, and the corresponding particle is copied into the resampled population. Particles with large weights may therefore be selected multiple times, whereas those with small weights may not be selected; the resulting population is then assigned equal weights.

Following resampling, each particle is rejuvenated using a No-U-Turn Sampler (NUTS; \citealt{NUTS}) transition that leaves $\pi_t$ invariant. This restores diversity lost from particles being resampled more than once and enables local exploration of the new tempered target. BlackJAX dynamically tunes the NUTS step size as the characteristic scale of the tempered target changes over the course of annealing. Repeating these stages until $\beta_T=1$ produces a particle approximation to the posterior.

For each fit, we used 4028 particles and applied one NUTS rejuvenation transition per particle at each inverse-temperature stage. We monitored the annealing schedule, particle ancestry, NUTS acceptance probability, trajectory length, and divergent transitions. For the complex-gain fit in the case of the primary reference tile we use throughout \S\ref{sec:results}, the sampler reached $\beta=1$ after 80 stages. The median NUTS trajectory length never exceeded 31 integration steps, and no trajectory reached the imposed ceiling of 1023 steps. Fewer than $0.08\%$ of particles diverged at any stage, and the final particle population retained both symmetry-related posterior modes.

\subsection{Spline model}

Here we discuss the spline model first in terms of its mathematical form and then in terms of its statistical structure.

\subsubsection{Spline representation}

We developed a Bayesian hierarchical model based on splines to parametrize deviations from the FEE beam, working in orthographic projection defined by the direction cosines,
\begin{align}
    & x = \sin(\theta)\cos(\phi) \\
    & y = \sin(\theta)\sin(\phi),
\end{align}
where $\theta$ is the polar angle from pointing center and $\phi$ is the azimuth. Since we only use zenith-pointed observations, $\theta$ is always the zenith angle. Splines are commonly used piecewise-continuous polynomials parametrized by a sequence of 'knots' (special locations) and the degree of the polynomials. The number of continuous derivatives at each knot is determined by how many times a knot is repeated in the sequence, as well as the degree of the spline. We use cubic splines, and only repeated knots at the boundaries, implying that the model has at least four continuous derivatives everywhere in orthographic projection. 

\begin{figure}
    \centering
    \includegraphics[width=\linewidth]{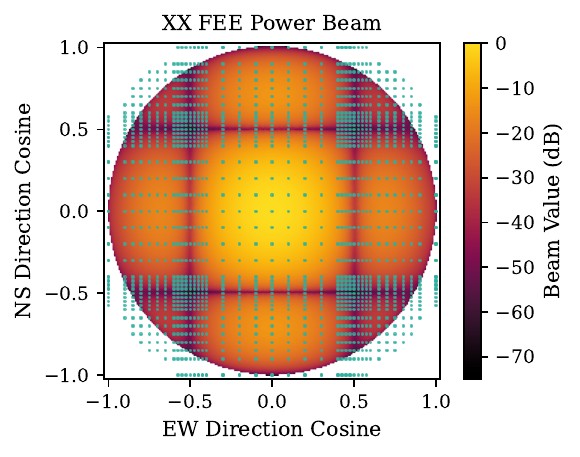}
    \caption{Orthographic projection of FEE beam at 137 MHz. In dB, it is mostly smooth except near the nulls. Superimposed green markers show how we have placed spline knots.}
    \label{fig:FEE}
\end{figure}

The motivation for splines was that the data naturally measure the logarithm of the beam (log-beam), and the log-beam exhibits localized scale structure. Near the nulls, the log-beam is quickly varying, but otherwise it is smooth (Figure \ref{fig:FEE}). By placing nonuniform spline knots, we can adjust the smoothness of the model in a localized fashion. Regions with denser knot placement can accommodate more rapid variation in the spline model, while regions of low knot density are more slowly varying. We hand-picked knots based on numerical experimentation with the FEE beam model. Our basic strategy was to divide the beam into three regions: the main lobe, the nulls, and the sidelobes. We picked the least dense knots in the main lobe, where the FEE model is smoothest, the densest knots near the nulls where the FEE model is sharply varying (in log), and a middle value for the sidelobes. We compared a few arbitrary prescriptions of this variety using an approximate Bayesian information criterion \citep[BIC, Appendix \ref{sec:BIC};][]{BIC} on an example tile. Finding these to score the best (cf. Appendix \ref{sec:BIC}), we left them fixed for all statistical inference presented here. A good knot selection for an arbitrary beam likely depends on the beam in question, so we encourage future users to explore their own models. However, in the absence of any clear choices, we reckon a sufficiently dense uniform placement of knots will suffice.

To perform Bayesian inference with the spline model, we took advantage of the B-spline representation. In this representation, a spline is a linear superposition of localized basis functions, whose can be calculated at arbitrary locations using the Cox-de Boor algorithm \citep{Cox1972TheNE, deboor1972calculating}. For a two-dimensional function like a beam pattern, we can take the products of pairs of one-dimensional basis elements to form each 2-dimensional basis element. That is, we can write a model of a two-dimensional functions as
\begin{equation}
    m(x, y) = \sum_{jk}c_{jk} f_j(x) g_k(y).
\end{equation}
where $c_{jk}$ are the spline coefficients and $f_j$ and $g_k$ are the corresponding basis functions. Conceptually, we can collect each basis function (i.e. each pair of 1d basis functions) at a predetermined list of locations (where the beam measurements took place) into a column vector, and arrange each column into a design matrix, $\mathbfss{A}$. We can then define a model vector, $\mathbfss{m}$ and coefficient vector $\mathbfss{c}$ and define
\begin{equation}
    \mathbfss{m} \equiv \mathbfss{A} \mathbfss{c}
\end{equation}
For a spline of degree $p$, only $p + 1$ basis functions contribute to any given point, resulting in a highly sparse design matrix. In this two-dimensional context, only $(p+1)^2 = 16$ basis functions contribute to each point. Therefore, despite the fact that our model has roughly $1500$ parameters, model evaluation is relatively fast. In theory, other types of localized basis function approaches (e.g. wavelets) could present many of the advantages that splines present here and would be suitable alternatives.

\subsubsection{Statistical Assumptions}

The work of \citet{Chokshi2021} suggest that the FEE beam model is a reasonable first approximation to the tile-by-tile response. For this reason, we use the spline model above to parametrize \textit{deviations} of any given tile's beam from the FEE model, rather than the entire beam pattern. Supposing again that the data, $\mathbfss{d}$, are corrupted by additive noise, $\mathbfss{n}$, we can write this model as
\begin{equation}
    \mathbfss{d} = \mathbfss{Ac} + \mathbfss{f} + \mathbfss{n},
\end{equation}
where $\mathbfss{f}$ is the FEE model evaluated at the locations of the data. As in the dipole model, we suppose that the noise is independently distributed according to the soft Laplace distribution (Equation \ref{eq:soft_laplace}), where the scales vary per pixel according to Equation \ref{eq:noise_full_gore}. In this formulation, the model is constraining the difference (in dB i.e. the log of the ratio) between the measurements and the FEE beam.

Rather than place a prior directly on the spline coefficients, we instead establish a hierarchical model based also on the scaled soft Laplace distribution and a related distribution called a skew logistic distribution with probability density
\begin{equation}
    P(z | s, \alpha) = g(z;s,\alpha) \equiv \frac{\alpha e^{-z/s}}{(1 + e^{-z/s})^{\alpha + 1}}.
\end{equation}
This distribution has asymmetric exponential tails which we empirically determined as a property of spline coefficients that fit our data well. Specifically, we suppose that given two unknown scale parameters, $s_c$ and $s_\Delta$, and a shape parameter, $\alpha$, the probability density function of $\mathbfss{c}$ is
\begin{equation}
    P(\mathbfss{c} | s_c, s_{\Delta c}, \alpha) = \prod_{j=1}^N g(c_j; s_c, \alpha) h(\Delta_j^x(\mathbfss{c}); s_\Delta)h(\Delta_j^y(\mathbfss{c}); s_\Delta)
\end{equation}
where $\Delta_j^x$ is the second-order accurate finite difference of the  spline coefficients along the $x$-direction for interior knots and first-order for exterior ones (i.e. the default output of \texttt{Numpy}'s gradient function). The second term is inspired by, but not exactly equivalent to, total variation minimization. This is meant to regularize the solution in large gaps so that the model tends to remain close to zero and/or flat in the absence of data, but with heavy tails in case a highly nonzero coefficient (or change in spline coefficients) helps fit the data appreciably. Letting the scales be free parameters lets the model ``self-regulate'' based on the statistics of well-fitting spline coefficients, rather than forcing the user to pick arbitrary regularization strengths. We use uniform priors on the scale parameters from 1 to 100, in units of dB. This is a very generous prior, implying that spline models are allowed to generate deviations from the FEE model by tens of orders of magnitude at the high end.\footnote{For example, if the scale parameter is 100 dB, then observing a spline coefficient of 200 dB would not be unlikely in a model of this size.} The use of a soft Laplace instead of Laplace distribution makes the regularization more compatible with gradient-based optimization and Monte Carlo techniques for a similar overall functionality. 

As shown in Appendix \ref{sec:hierarchical_structure}, the soft Laplace is a reasonably accurate shape for the marginal distribution of the inferred finite differences. The skew logistic distributions somewhat fails to capture the marginal distribution of the inferred coefficients themselves, but the heavy tails theoretically prevent excessive ``shrinking'' that might arise from e.g. a Gaussian-shaped prior with thinner tails. Nonetheless, the model seems to broadly fit the data and produce consistent results over various jackknives, so we consider this inaccuracy acceptable grounds for future work. This hierarchical structure may be unnecessary when the beam pattern is sampled with sufficiently fine resolution (and sufficiently low noise) as in \S\ref{sec:full_data}, since the data alone will provide enough information to determine the coefficients. However, the structure is of enormous benefit when the beam pattern is only sparsely sampled, as we demonstrate in \S\ref{sec:time_jackknife}.

We implement Bayesian inference for this model using the \texttt{Numpyro}\footnote{\url{https://num.pyro.ai}} probabilistic programming library. We make use of the NUTS algorithm implemented therein to generate posterior samples. For any given inference, we generated four independent chains and tuned the warm up and sampling lengths until we obtained effective sample sizes exceeding 1000 and Gelman-Rubin statistics $\hat{R} \lesssim 1.01$ for all parameters. Collecting 6000 samples per chain with the first 2000 as warmup generated effective sample sizes typically in the several thousands for well-behaved data subsets. Each chain takes roughly 2 minutes using an Apple M3 Pro, independent chains can be parallelized, and this figure is faster for smaller data sets. We process all tiles and polarizations independently, producing posterior samples for each.

\section{Results}
\label{sec:results}

We present detailed results for the S06XX power beam, where S06 denotes tile 06 in the southern hexagon of the MWA Phase~II compact array and XX denotes the response of its east--west-aligned dipoles~\citep{Chokshi2021}. We first use the full dataset to identify coherent departures from the nominal AEE and FEE beams and determine how much of this structure is captured by the physically motivated dipole-gain model and the hierarchical spline model. We then repeat both inferences using independent 20-day subsets to test which features remain reproducible under sparse and non-identical sampling patterns. Corresponding results for S10YY are presented in Appendix~\ref{sec:ant_10_demo}.

\subsection{Full Data Set}
\label{sec:full_data}

\begin{figure}
    \centering
    \includegraphics[width=\linewidth]{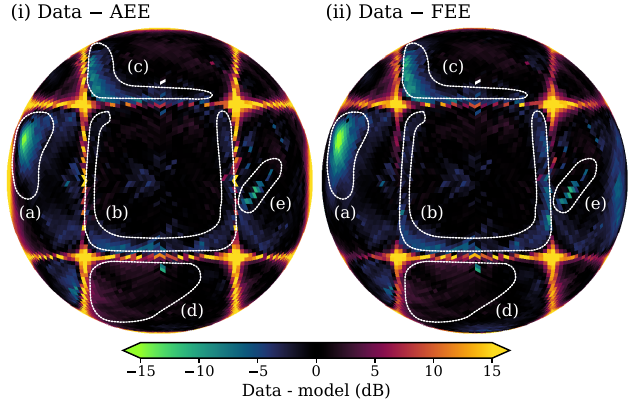}
    \caption{Residuals of the S06XX satellite measurements relative to the nominal (i) AEE and (ii) FEE beam models, evaluated on the same HEALPix grid. The prominent orange hash-shaped structure traces the large positive residuals at the nominal beam nulls. Identical dotted outlines mark (a) the western-sidelobe depression, (b) the partial-square residual along the inner null boundaries, (c) and (d) opposing gradients across the northern and southern sidelobes, and (e) an isolated pass-shaped artefact. The outlines are included as visual guides only. Positive values indicate that the measured response exceeds the model prediction, and the shared colour scale is clipped at $\pm15~\mathrm{dB}$.}
    \label{fig:data_aee_fee_residuals}
\end{figure}

Before considering either inference model, we compare the measured S06XX beam directly with the nominal electromagnetic predictions. Figure~\ref{fig:data_aee_fee_residuals} shows the measurement residuals relative to the AEE and FEE models. Although the two electromagnetic models differ in detail, their residuals exhibit the same principal spatial features.

The most prominent discrepancy occurs along the nominal beam nulls, visible as the bright orange hash-shaped pattern in Figure~\ref{fig:data_aee_fee_residuals}. The satellite measurements span a dynamic range exceeding $60~\mathrm{dB}$, while the deepest nulls in the AEE and FEE predictions reach approximately $-50~\mathrm{dB}$ (Figure~\ref{fig:aee_fee_comparison}). Nevertheless, the measured response within these regions generally remains above approximately $-30~\mathrm{dB}$. Comparably shallow nulls were observed in all 14 tiles measured by \citet{Chokshi2021}. The shallower nulls therefore represent a systematic discrepancy between the measured tile responses and the electromagnetic predictions, rather than a simple limitation of measurement sensitivity. This discrepancy is particularly consequential for high-dynamic-range experiments such as measurements of the redshifted 21-cm signal. Errors of $10$--$30~\mathrm{dB}$ in the beam response correspond to errors of one to three orders of magnitude in the predicted apparent power of a source passing through a null. For sufficiently bright sources, such beam mismodelling could therefore introduce substantial errors into calibration procedures that rely on forward modelling the apparent sky~\citep{EwallWice2017,Chokshi2024}.

Narrow residual features also trace the inner boundaries of the nulls, forming a partial square around the main lobe (Figure~\ref{fig:data_aee_fee_residuals}, region~(b)). Their spatially coherent morphology is consistent with the measured nulls being slightly displaced relative to both nominal predictions. The residual maps also reveal coherent structure away from the nulls. The western sidelobe contains a broad depression in which the measured response is systematically weaker than predicted (region~(a)), while the northern and southern sidelobes exhibit opposing large-scale gradients (regions~(c) and (d)). Alternating-sign residual patterns are also visible across both the cardinal and diagonal sidelobes. These sidelobe structures extend smoothly across contiguous regions and differ morphologically from the narrow pass-shaped artefact highlighted in region~(e). Together, these shared residual features define the empirical structure that the two inference models must explain. We first use the dipole model to determine how much can be represented by physically interpretable variations in the dipole gains, and then use the spline model to identify additional structure supported by the measurements.

To determine how much of the observed structure can be represented by the physically motivated dipole model, we first fit a reduced amplitude-only form in which all dipole phases are fixed to zero and only the 15 independent amplitudes are inferred. We then fit the full complex form, allowing the 15 relative phases to vary in addition to the amplitudes. Table~\ref{tab:model_comparison} summarises the resulting residual statistics and model-comparison measures. 

\begin{table}
    \centering
    \begin{tabular}{@{}lrrrrr@{}}
        \hline
        Model & $k$ & RMS (dB) & MAD (dB) & BIC & $\log Z$ \\
        \hline
        Nominal   & 0  & 7.889 & 1.749 & $4.04\times10^{6}$ & -- \\
        Amplitude & 15 & 2.945 & 1.188 & $8.22\times10^{4}$ & $-41132.1$ \\
        Complex   & 30 & 2.890 & 1.015 & $7.44\times10^{4}$ & $-37214.9$ \\
        \hline
    \end{tabular}
    \caption{Comparison of the nominal, the amplitude-only and complex-gain forms of the dipole model for S06XX. Here, $k$ is the number of inferred parameters. The RMS and median absolute deviation (MAD) are evaluated over the likelihood pixels using either the nominal beam or the posterior-weighted mean inferred beam. For the complex model, the beam statistics use the more probable of the two equivalent gain configurations discussed below, while the Bayesian evidence includes the complete posterior.}
    \label{tab:model_comparison}
\end{table}

Allowing the dipole amplitudes to vary removes most of the discrepancy between the measurements and the nominal AEE beam. The amplitude-only model reduces the residual RMS from $7.889$ to $2.945~\mathrm{dB}$ and the MAD from $1.749$ to $1.188~\mathrm{dB}$. Allowing the gains to be complex produces a smaller additional reduction in RMS, from $2.945$ to $2.890~\mathrm{dB}$, but reduces the MAD more appreciably, from $1.188$ to $1.015~\mathrm{dB}$.

The additional phase parameters are strongly preferred after accounting for the greater model complexity. Relative to the amplitude-only model, the complex model lowers the BIC by $7.84\times10^{3}$ and increases the log Bayesian evidence by $\Delta\log Z =\log Z_{\mathrm{complex}}-\log Z_{\mathrm{amplitude}} =3917.2$. Both measures therefore strongly favour the complex-gain model.

We next examine the inferred gains and the different symmetry structures of the amplitude-only and complex parametrisations. In the amplitude-only fit, the posterior median amplitudes span $0.68$--$1.40$, representing substantial departures from the nominal value of unity. Restricting the gains to positive real values also changes the symmetry structure of the permitted parameter space. The conjugate-rotation transformation derived in \S\ref{sssec:symmetry} maps the effective coefficients of the amplitude-only solution onto a second configuration with an identical power beam. The transformed gains generally have non-zero phases, and therefore lie within the full complex-gain parameter space but outside the amplitude-only model. Consequently, only one member of the symmetry-related pair is present in the amplitude-only posterior.

In the full complex-gain model, we also allow the 15 relative dipole phases to vary under a von Mises prior. In this enlarged parameter space, the posterior contains two clearly separated modes related by the conjugate-rotation symmetry. We classify the two modes using a normalised symmetry-branch statistic
\begin{equation}
    S_p
    =
        \sum_{n=0}^{7}
        \left(
            \lvert c_{n,p}\rvert^2
            -
            \lvert c_{15-n,p}\rvert^2
        \right)
    \bigg/
        \sum_{n=0}^{15}
        \lvert c_{n,p}\rvert^2.
    \label{eq:symmetry_statistic}
\end{equation}
The numerator compares the total coefficient power between the two members of each dipole pair related by the $180^\circ$ tile rotation, while the denominator removes any dependence on their common scale. Under conjugate rotation, the paired coefficient amplitudes are exchanged, and hence $S'_p=-S_p.$ The sign of $S_p$ therefore provides a deterministic label for the two symmetry-related modes. The labels $S_p>0$ and $S_p<0$ identify their orientations only and have no independent physical significance.

\begin{figure}
    \centering
    \includegraphics[width=\columnwidth]
        {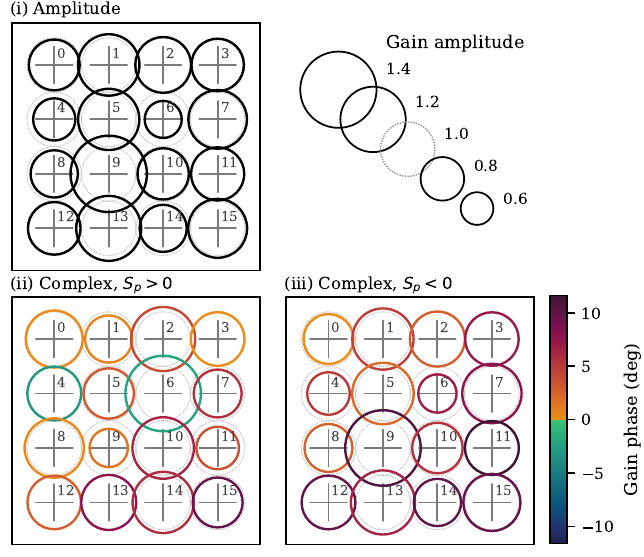}
    \caption{Inferred dipole gains for the full S06XX dataset. Panel~(i) shows the amplitude-only solution, while panels~(ii) and (iii) show the $S_p>0$ and $S_p<0$ modes of the complex-gain posterior, respectively. The radius of each solid circle represents the dipole-wise posterior median amplitude, with the dotted circle marking the nominal unit amplitude. For the complex solutions, the circle colour represents the posterior circular-mean phase relative to the fixed reference $\phi_{0,p}=0$. The amplitude-only solution agrees with the $S_p<0$ complex solution, and with the $180^\circ$-rotated $S_p>0$ solution, to an RMS difference of $0.016$ in linear gain amplitude.}
    \label{fig:gain_solutions}
\end{figure}

Within the complex-gain fit, the final SMC population assigns $93.3\%$ of its posterior mass to the $S_p>0$ mode and $6.7\%$ to the $S_p<0$ mode. Figure~\ref{fig:gain_solutions} shows the amplitude-only solution and the branch-conditioned amplitude and phase estimates from the complex-gain model. Their ranges and posterior masses are summarised in Table~\ref{tab:dipole_gain_summary}.

\begin{table}
    \centering
    \caption{Summary of the inferred dipole gains for the full S06XX dataset. Amplitude ranges are taken across the dipole-wise posterior medians. Phase ranges are taken across the posterior circular means of the 15 inferred relative phases, excluding the fixed reference phase $\phi_{0,p}=0$. The posterior masses of the complex modes are calculated from the final SMC particle weights.}
    \label{tab:dipole_gain_summary}
    \begin{tabular}{llccc}
        \hline
        Model & Mode & Amp. range & Phase range & Posterior mass \\
        \hline
        Amplitude
            & -- & $0.68$--$1.40$ & $0^\circ$ (fixed) & -- \\
        Complex
            & $S_p>0$ & $0.70$--$1.39$
            & $-3.1^\circ$--$9.7^\circ$ & $93.3\%$ \\
        Complex
            & $S_p<0$ & $0.70$--$1.39$
            & $2.1^\circ$--$11.6^\circ$ & $6.7\%$ \\
        \hline
    \end{tabular}
\end{table}

The inferred amplitude structure is remarkably consistent between the two models. The amplitude-only posterior medians agree with those of the $S_p<0$ complex mode, and with those of the $S_p>0$ mode after a $180^\circ$ rotation, to an RMS difference of $0.016$ in linear gain amplitude. Introducing the phase degrees of freedom therefore leaves the inferred amplitude pattern nearly unchanged, apart from the rotation relating the two complex modes.

\begin{figure}
    \centering
    \includegraphics[width=\columnwidth]{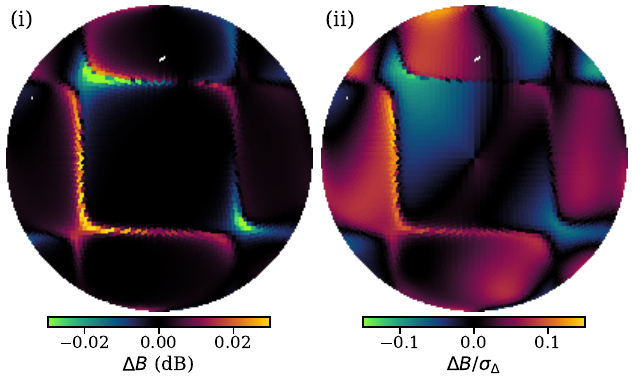}
    \caption{Comparison of the power beams inferred from the two symmetry-related modes of the S06XX complex-gain posterior. Let $\overline{B}_{+}$ and $\overline{B}_{-}$ denote the weighted mean beams calculated from particles in the $S_p>0$ and $S_p<0$ modes, respectively, with corresponding posterior standard deviations $\sigma_{+}$ and $\sigma_{-}$. Panel~(i) shows their difference, $\Delta B=\overline{B}_{+}-\overline{B}_{-}$, while panel~(ii) shows this difference relative to $\sigma_\Delta=(\sigma_{+}^{2}+\sigma_{-}^{2})^{1/2}$. The mean beams differ by only $0.0056~\mathrm{dB}$ RMS, with $\lvert\Delta B\rvert/\sigma_\Delta<0.13$ throughout the evaluated region.}
    \label{fig:complex_mode_comparison}
\end{figure}

Despite their different gain configurations, the two complex modes predict the same power beam. Applying the conjugate-rotation transformation to individual posterior particles changes the predicted beam by only approximately $4\times10^{-6}~\mathrm{dB}$ RMS, consistent with numerical precision. For a population-level comparison, we divided the posterior particles according to the sign of $S_p$ and calculated a weighted mean beam within each mode. Figure~\ref{fig:complex_mode_comparison} compares these mean beams, which differ by only $0.0056~\mathrm{dB}$ RMS. Relative to the quadrature sum of their within-mode posterior uncertainties, the difference is less than $0.13\sigma$ in every evaluated pixel. The distinct gain configurations therefore represent the same inferred power beam.

Because the two modes have indistinguishable likelihoods, their unequal posterior masses arise primarily from the phase prior. The $S_p>0$ solution has smaller overall circular phase departures from zero and is therefore assigned greater density by the zero-centred von Mises prior. For a solution near the peak of the preferred mode, the RMS phase displacement is $4.64^\circ$, compared with $7.45^\circ$ for its exact symmetry partner. Their phase-prior density ratio corresponds to relative weights of approximately $92.7\%$ and $7.3\%$, closely matching the SMC posterior masses. The imbalance is therefore predominantly induced by the phase prior, which favours the representation closer to the nominal zero-phase configuration, as intended.

\begin{figure*}
    \centering
    \includegraphics[width=\linewidth]{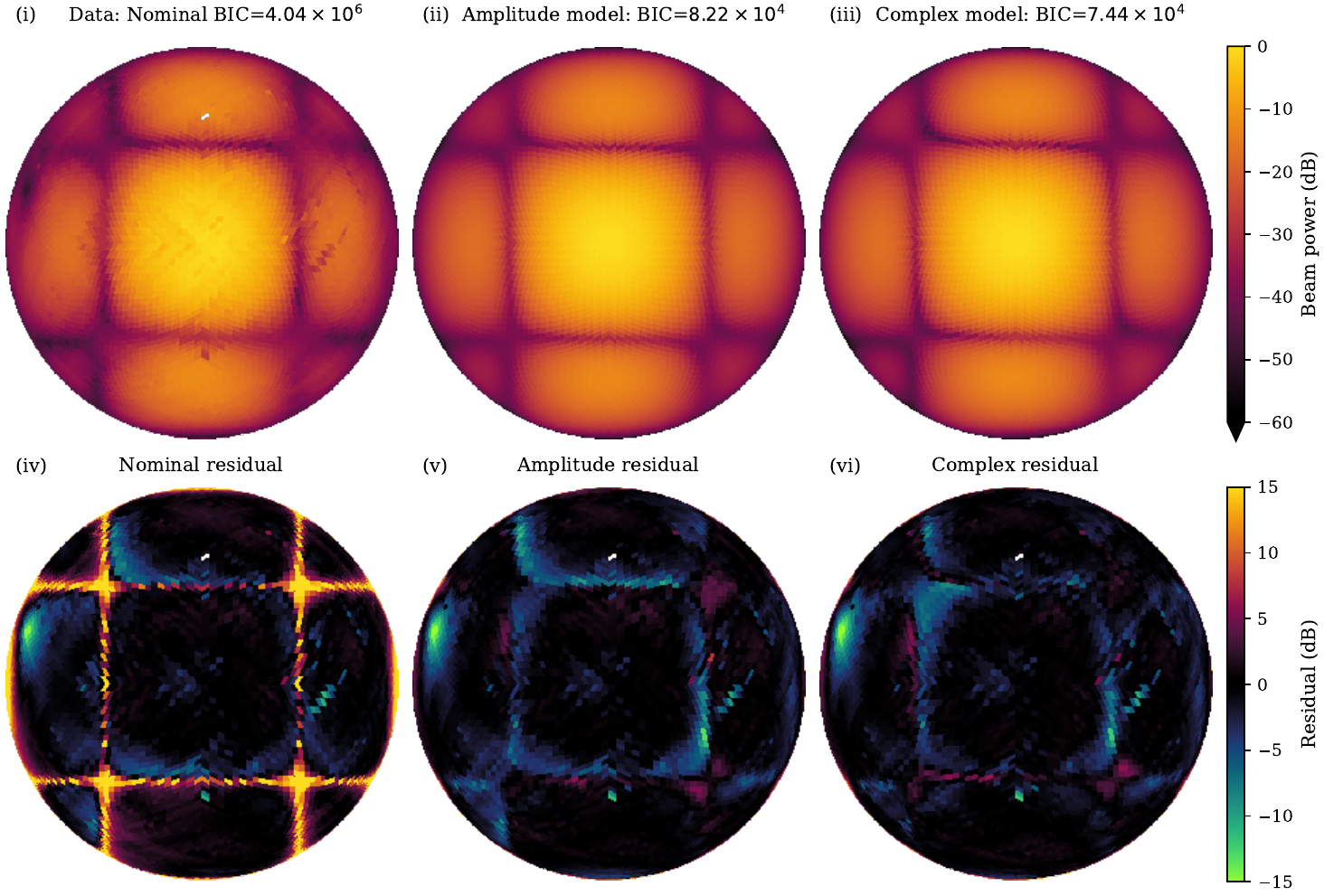}
    \caption{Measured and inferred S06XX beam responses. Panel~(i) shows the satellite measurements. Panels~(ii) and (iii) show the posterior-weighted mean beams inferred using the amplitude-only and complex-gain models, respectively; the latter is calculated from the dominant $S_p>0$ mode. Panels~(iv)--(vi) show the residuals $r_q=d_q-m_q$ relative to the nominal AEE, amplitude-only, and complex-gain models. Allowing the dipole gains to vary removes much of the discrepancy with the nominal beam, while the complex-gain model produces a further reduction in the residual structure.}
    \label{fig:dipole_beam_comparison}
\end{figure*}

\begin{figure}
    \centering
    \includegraphics[width=\columnwidth]{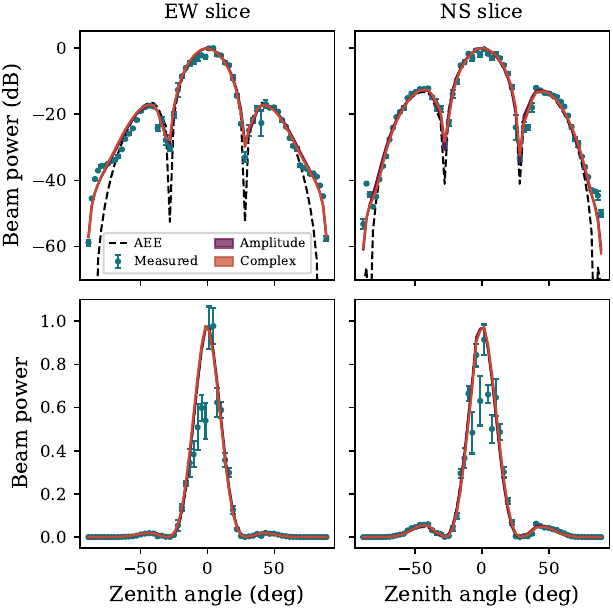}
    \caption{Approximate east--west (left) and north--south (right) slices through the S06XX beam in logarithmic (top) and linear (bottom) power. The dashed curves show the nominal AEE beam, while the coloured curves and shaded regions show the posterior-weighted mean and one-standard-deviation uncertainty for the amplitude-only and complex-gain dipole models. The complex-gain result is conditioned on the dominant $S_p>0$ posterior mode. Points show the measured beam values, with error bars representing the corresponding noise estimates. Negative zenith angles are West or South.}
    \label{fig:dipole_beam_slices}
\end{figure}

For the following beam comparisons, we use the dominant $S_p>0$ mode as the representative complex solution. We condition the posterior on this mode by retaining only the particles assigned to it and rescaling their SMC weights to sum to unity. The complex-model beam and its residual statistics are then calculated using this conditional posterior. This avoids averaging between distinct gain configurations while having a negligible effect on the predicted power beam. Figure~\ref{fig:dipole_beam_comparison} compares the measured beam and spatial residuals of the nominal AEE, amplitude-only, and complex-gain models, while Figure~\ref{fig:dipole_beam_slices} shows their east--west and north--south beam profiles.

The beam maps and slices show that the dipole model captures much of the broad departure from the nominal AEE, consistent with the substantial reductions in RMS and MAD. The complex model provides the statistically preferred physical description, but neither form produces noise-like residuals. Spatially coherent discrepancies remain around the nulls and throughout the sidelobes, including structures associated with the regions identified in Figure~\ref{fig:data_aee_fee_residuals}. The inferred gains therefore provide a compact physical representation of much of the measured deformation, but variations in dipole gains are insufficient to describe the measured beam completely. This remaining structure motivates consideration of departures from the fixed nominal tile geometry before we turn to the more spatially flexible spline model.

Although the dipole model is physically motivated, the inferred gains should be interpreted as effective parameters rather than unique measurements of electronic gain errors. The model holds the dipole positions, orientations, embedded-element responses and ground-screen geometry fixed at their nominal values. Variations in the complex gains can describe genuine excitation differences and may absorb part of the response to other perturbations, but their success does not necessarily establish the physical origin of the measured deformation. For example, a displacement $\delta\mathbf{r}_n$ of the $n$th dipole introduces a phase factor $\exp(i k\delta\mathbf{r}_n\mathbin{\cdot}\hat{\mathbf{s}})$ that varies across the sky. A single direction-independent complex gain may approximate this effect over a limited region, but cannot generally reproduce it over the full hemisphere.

Departures from the nominal tile geometry provide plausible explanations for several features in Figure~\ref{fig:data_aee_fee_residuals}. The perfectly regular and coplanar dipole grid assumed by the AEE and FEE models produces sharp, deep nulls through coherent destructive interference. Small errors in dipole position, height or orientation would perturb the relative phases across the tile, reducing this coherence and making the nulls shallower. A nearly rigid tilt of the tile or ground screen would instead displace the beam centre and could produce the partial-square residual along the inner null boundaries in region~(b). A small azimuthal rotation away from the assumed cardinal alignment could also account for at least part of the opposing gradients in the northern and southern sidelobes, highlighted by regions~(c) and (d), as well as the alternating-sign residuals across the cardinal and diagonal sidelobes. Such a rotation naturally produces residuals of opposite sign on opposing flanks of the nominal beam features.

The broad western depression in region~(a) is less readily explained by a single global tilt or rotation and may instead require a more localised departure from the nominal geometry. Warping of the ground screen, asymmetric erosion, accumulated soil or vegetation growth could modify the response in particular directions. For example, \citet{Chokshi2021} documented large rocks along the western edge of tile S08's ground screen and loose soil covering parts of its conducting mesh, illustrating the extent to which installed tiles can depart from their idealised electromagnetic geometry. We cannot attribute the S06XX residuals to any specific mechanism from these measurements alone, and later use the 20-day jackknives to test their reproducibility rather than their cause. We therefore turn to the spline model to determine which additional direction-dependent structure is supported by the data without requiring it to arise from a particular physical mechanism.

\begin{figure*}
    \centering
    \includegraphics[width=\linewidth]{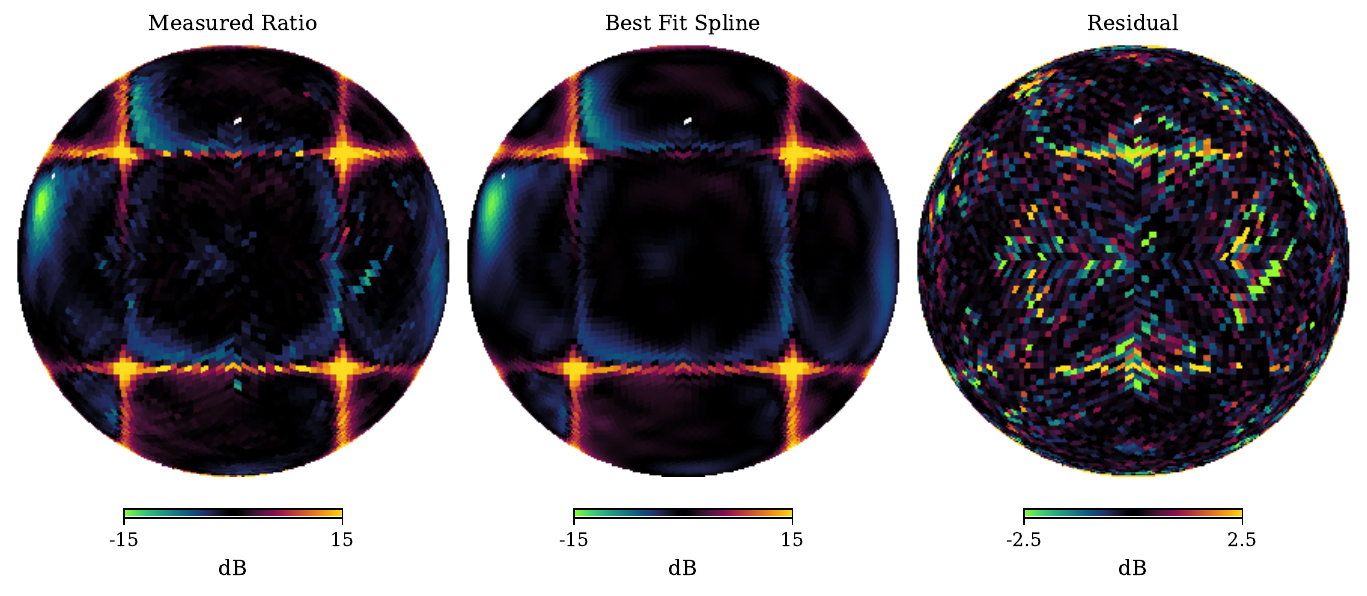}
    \caption{Comparison of measurements to model for an example tile (06, XX polarization), using the full data set. White pixels had fewer than 3 total contributing data points. The spline model (middle) mainly extracts smooth structure from the ratio of the measured beam to the FEE model (left), avoiding some pass-shaped artefacts while capturing the sharp structure near the nulls effectively. Residuals demonstrate pass-shaped artefacts, as well as systematic errors in the East-West aligned nulls.}
    \label{fig:spline_example}
\end{figure*}

The spline model provides a complementary, spatially flexible reconstruction of the measured beam. A principal result of this analysis is that it recovers structure over a range of angular scales without being dominated by pass-shaped artefacts in the binned HEALPix maps. Figure~\ref{fig:spline_example} shows the best-fitting spline correction to the ratio of the binned S06XX beam to the nominal FEE model evaluated at the HEALPix pixel centres. The measured ratio contains fine-scale structure near the beam nulls and smoother structure throughout the sidelobes. Visible pass-shaped outliers, such as the feature highlighted by region~(e) in Figure~\ref{fig:data_aee_fee_residuals}, are not reproduced by the spline and instead remain apparent in its residuals.

There is also a coherent depression of the measured beam for this tile in the Western sidelobe as well as at the horizon. This particular feature seems to exist in more than one tile, but varies from tile-to-tile and its location is polarization-dependent (Appendix \ref{sec:ant_10_demo}). While some random blobs recovered by the spline appear to align with specific outlier passes, this feature appears different in that it doesn't follow an obvious arcing pattern. We are unsure what exactly could cause a depression like this in the sidelobe, especially since it is not particularly well-captured by the modified AEE model. However, a beam model that can accommodate some features like this will still be useful for precision experiments relying on accurate beam models. 

\begin{figure*}
    \centering
    \includegraphics[width=\linewidth]{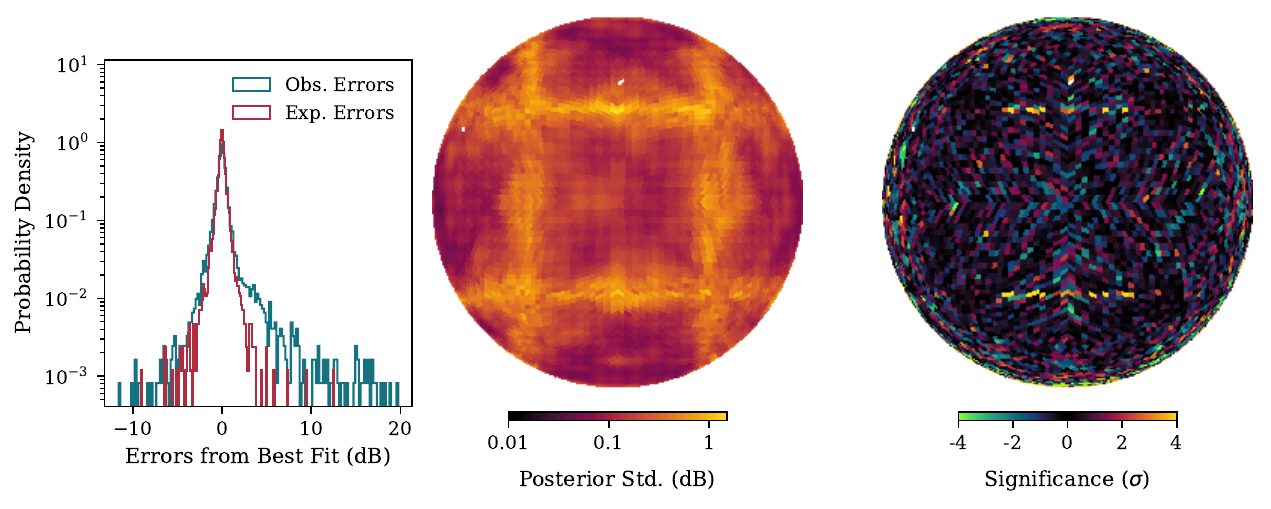}
    \caption{Left: observed errors with respect to best-fit spline, along with samples from the noise distribution for reference (expected errors). Middle: Posterior uncertainty in the spline fit. Right: Significance, mapped to normal standard deviations, of data with respect to posterior predictive distribution of the spline model (i.e. the statistical significance of the residuals \textit{including} posterior uncertainty). }
    \label{fig:error_stats}
\end{figure*}

The residuals demonstrate that pass-shaped artefacts in the measured ratio (region~(e), Figure~\ref{fig:data_aee_fee_residuals}) are generally ignored by the spline when they appear as single arcs, but if multiple arcs appear together with the same sign, they will generate a smooth blob in the spline model. While we find this association suspicious, it is true that multiple nearby arcs appearing with the same sign and similar magnitude represents a genuine statistical consistency of the data and therefore something that the model ought to fit. We also noticed that with a Gaussian noise model, rather than soft Laplace, these blobs are much more common, filling the inferred beam with smooth ripples (not shown). This demonstrates that the soft Laplace noise model helps make the the inference more robust against these outliers, along with accurately describing the error statistics (Figure \ref{fig:error_stats}, left subplot). We are surprised that these arcs appear in the measurements at all, since with the full data set typically many passes combine for each pixel, especially close to the horizon. We suspect that certain passes are significantly miscalibrated compared to others despite extensive efforts to exclude poorly calibrated data from the analysis through various means, leading to biased pixels. This hypothesis and the nature of the smooth coherent blob in the Western sidelobe are addressed in our time jackknife (\S\ref{sec:time_jackknife}).

The shape of the noise distribution is seen to be accurate by comparing the difference between the best-fit spline and the measurements to the expected noise distribution. The left subplot of Figure \ref{fig:error_stats} shows this by computing monte carlo samples from the soft Laplace distribution corresponding to the error estimate at each HEALPix pixel from Equation \ref{eq:noise_full_gore}. The monte carlo samples (orange) have an extremely similar distribution as the actual residuals (blue) with the exception of a small set of positive outliers. These positive outliers account for roughly five percent of the data, and are almost exclusively located near the horizon at zenith angles greater than $85^\circ$. 

Two key assumptions went into the noise model: a marginal shape, checked by the analysis in the preceding paragraph, and the assumption that the noise in each pixel is independently distributed. It is easier to see that this second assumption is violated if we map the observed data to their statistical significance with respect to the posterior predictive distribution, which is the distribution of hypothetical future experiments conditional on having observed this data. To generate samples from this distribution, we add soft Laplace samples with appropriate scale parameter to each posterior sample evaluated at each pixel. We can then take the quantile of the data with respect to these samples and map them back to $z-$scores using the standard normal inverse cumulative distribution function. This gives a measure of statistical significance that takes into account the full posterior uncertainty in the model (Figure \ref{fig:error_stats}, middle subplot) as well as the heteroscedasticity of the noise statistics.\footnote{Here, the posterior uncertainty seems to track knot density rather than the noise map (Figure \ref{fig:noise_map}). We observe fairly strong degeneracies in inferred coefficients for nearby spline basis functions (correlation coefficients up to about 0.8), and suspect this is the cause of this pattern.} This mapping (Figure \ref{fig:error_stats}, right subplot) teases out correlated structure in the residuals that we believe represent lingering systematics in the data.

There are two primary features we notice in the significance map. First, there are clear correlated structures in the form of diagonally separated pixels of similar color. These diagonal structures are most noticeable in the main lobe, where fewer passes occurred, and resemble the pass-shaped artefact highlighted by region~(e) in Figure~\ref{fig:data_aee_fee_residuals}. We suspect that they are biases from particular passes. The fact that they show up in this map, which we can think of as an uncertainty-weighted residual map, means that that the spline model is avoiding the structures to a large degree. Meanwhile, residuals in the sidelobe tend to appear more random, though not entirely so. Importantly notice that the region of the coherent depression of the beam pattern in the Western sidelobe is not inhabited by a collection of similarly-signed residuals, indicating it is probably not a random blob from pass biases. The second feature we point out is that the residuals near the EW-aligned nulls do appear highly statistically significant, indicating that there may be a failure of the model to exactly capture the structure of the measured ratio near the nulls. Examining the measured ratio in that location, this would require a somewhat oscillatory spline solution, which is likely downweighted by the regularization scheme (i.e. hierarchical structure of the model). We find it unlikely that is the result of two pass-based biases happening to land exactly along the nulls, but are puzzled over the physical meaning of such a residual.

One possibility for this oscillatory feature, which we point out exists in other tiles (Appendix \ref{sec:ant_10_demo}), is mutual coupling between different tiles of the MWA.\footnote{Note this is coupling between full tiles, not the coupling handled by the FEE model between dipoles within a single tile.} Reradiated, reflected, or otherwise scattered radiation from other tiles inject copies of the measured signal into a given tile. Based on the baseline between tiles, this can imprint a fringing structure in the beam pattern. Tile-tile coupling generally appears in the sidelobes, and can produce rather unpredictable structure depending on the array \citet{Gueuning2026}. Mutual coupling in the MWA Phase II compact array was explicitly studied by \citet{Elder2024}. Coupling strengths were generally found to be small -- generally less than -40 dB at this frequency depending on the tiles -- which is roughly the strength of this feature.

\begin{figure}
    \centering
    \includegraphics[width=\linewidth]{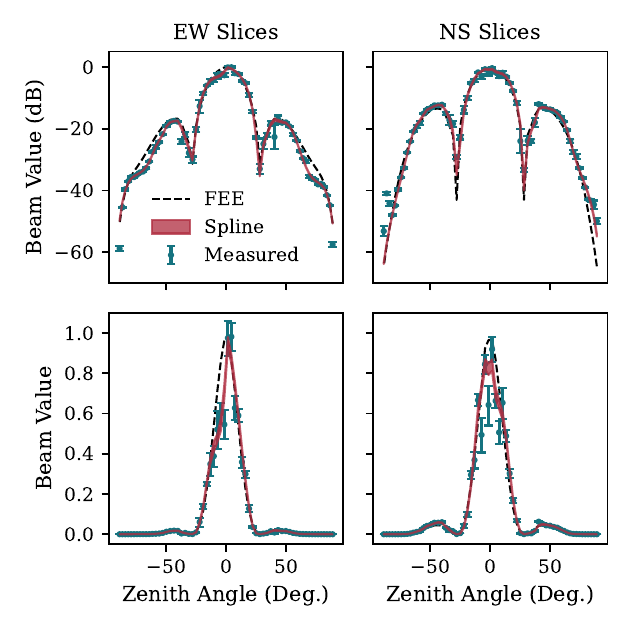}
    \caption{Approximate East-West (left) and North-South (right) slices through the beams of tile S06XX in logarithmic (top) and linear (bottom) scales. Errorbars and shaded regions represent a single standard deviation from Monte Carlo sampling or analytic noise estimation as appropriate. Negative zenith angles are West or South.}
    \label{fig:spline_slice}
\end{figure}

We examine slices through the measured beam and its models in Figure \ref{fig:spline_slice}. These slices are approximately East-West aligned and North-South aligned to the extent possible with \texttt{NSIDE}~$=32$.\footnote{Note that this is a restriction of the binning procedure whereas the models can be evaluated anywhere on the upper hemisphere.} Overall, the spline model is an excellent fit to the data, generally maintaining smoothness while being robust to the occasional strange outlier (see e.g. near the horizon in the upper-right panel). However, the data have the peculiar feature that there is a noticeable depression in the main lobe. 

The main lobe of the MWA primary beam is an essential ingredient to many scientific results made with the instrument, and we suspect deviations as large as what we see here would have made a noticeable impact on the consistency of results from the instrument. Furthermore, the main lobe residuals in Figures \ref{fig:spline_example} and \ref{fig:error_stats} exhibit obvious pass-shaped structures, and passes through the main lobe were known to cause performance issues for the RF Explorers. Note that before the recalibration procedure described in \S\ref{sec:calibration}, measurements (and hence the inferred spline) were at values less than 0.5 in linear units (i.e. biased low by a few dB). We managed to recover much of the main lobe shape with the recalibration procedure, but believe the data to possess systematic effects that are captured by the spline. We could prevent this by incorporating a separate prior for spline basis functions in the main lobe, which would shift the pattern to appearing the residuals. As a general demonstration of the spline model, we see that it can accommodate smooth deformations wherever they appear in the data, and are not particularly troubled by this. 

Taken together, the dipole and spline models provide complementary descriptions of the measured beam. The dipole model captures much of the broad departure from the nominal AEE using a compact and physically interpretable parametrisation, while the spline model provides a more complete phenomenological reconstruction of the localised null and sidelobe structure. Features recovered by both models are insensitive to their substantially different parametrisations, whereas structure captured only by the spline demonstrates the limitations of direction-independent dipole gains applied to a fixed nominal geometry. The spline model can, however, reproduce any sufficiently smooth structure consistently supported by the measurements, irrespective of whether it originates in the beam or in a persistent systematic effect. We therefore repeat both inferences using independent 20-day subsets to determine which features remain reproducible under sparse and non-identical sampling patterns.

\subsection{Time Jackknife}
\label{sec:time_jackknife}

Because both beam models describe spatially smooth patterns, they can recover beam structure even with limited coverage of the hemisphere. We test the stability of these reconstructions by dividing the first 80 days of observations into four non-overlapping 20-day chunks and repeating the inference independently for each subset. Each chunk contains approximately $20\%$ of the measurements used in the full-data analysis; an additional, more sparsely distributed set of observations obtained after these 80 days is not included in the jackknife. Since the four chunks contain different satellite passes, they provide distinct and incomplete sampling patterns across the hemisphere. Comparing their inferred beams allows us to identify spatial features that remain reproducible under changes in temporal coverage and sampling geometry, and to distinguish them from features that depend strongly on particular passes or subsets of the data.

\begin{figure}
    \centering
    \includegraphics[width=\linewidth]{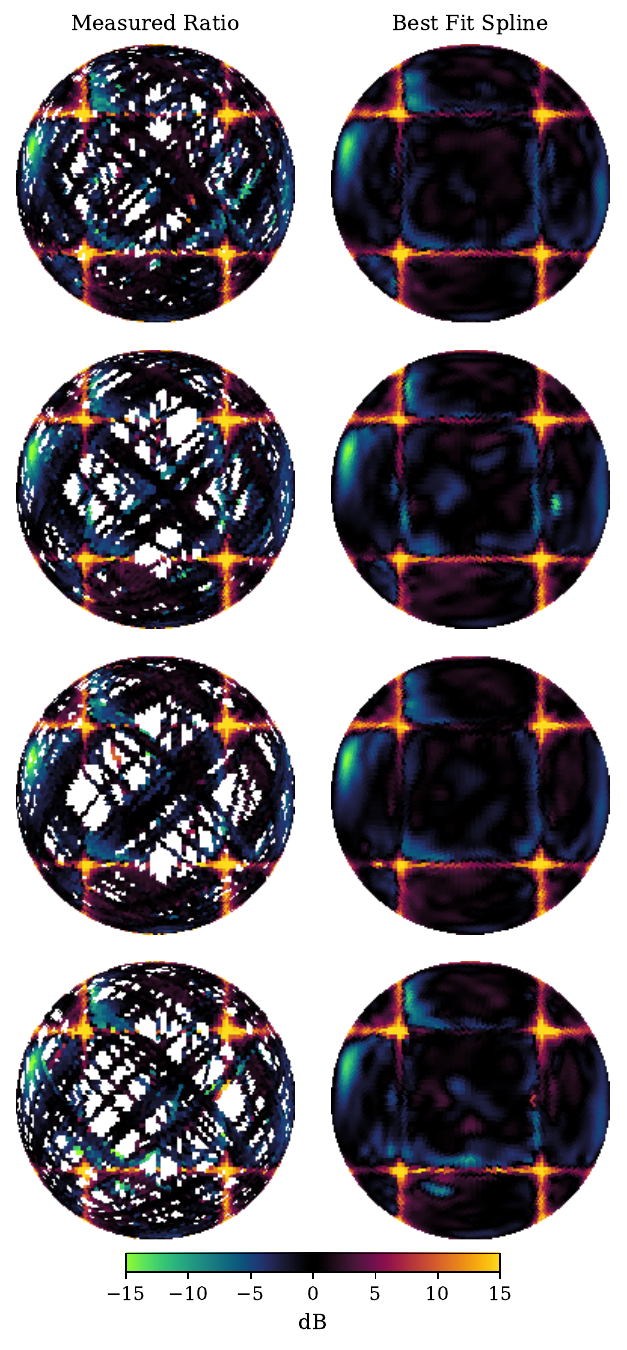}
    \caption{Left: Measured ratios to the FEE model for four 20-day chunks of data. White points are unsampled. Right: Corresponding best-fit splines, including evaluations at unsampled points.}
    \label{fig:chunk_fits}
\end{figure}

Sampling is particularly limited in the main lobe. Satellites passing near zenith generally have larger apparent angular velocities, so fewer measurements are obtained there and each pixel is typically supported by fewer independent passes. This increases both the noise of the binned measurements and their susceptibility to produce individual pass-dependent effects in the inferred beams. We fit each 20-day chunk independently with both the spline and dipole models, beginning with the spline model. The left column of Figure~\ref{fig:chunk_fits} shows the measured satellite-to-FEE beam ratio for each chunk, making their distinct sampling patterns explicit, while the right column shows the corresponding inferred beam ratios including unsampled locations on the hemisphere.

Despite their distinct and relatively sparse sampling patterns, the null and sidelobe structure of the best-fitting spline reconstructions is highly consistent across the four chunks. Each 20-day chunk supports the shallower and displaced nulls observed in the full-data inference, as well as the large coherent depression in the western sidelobe. They also all recover the weaker depression in the far eastern sidelobe. However, the main lobe and some sidelobes contain apparently spurious structures (random blobs) that appear in some time chunks but not others. As expected from the particularly sparse main-lobe coverage, these subset-dependent features are most prominent where individual passes have greater influence on the inference. Overall, it is encouraging that the principal null and sidelobe structure can be recovered consistently using only $\sim20\%$ of the data.

\begin{figure}
    \centering
    \includegraphics[width=\linewidth]{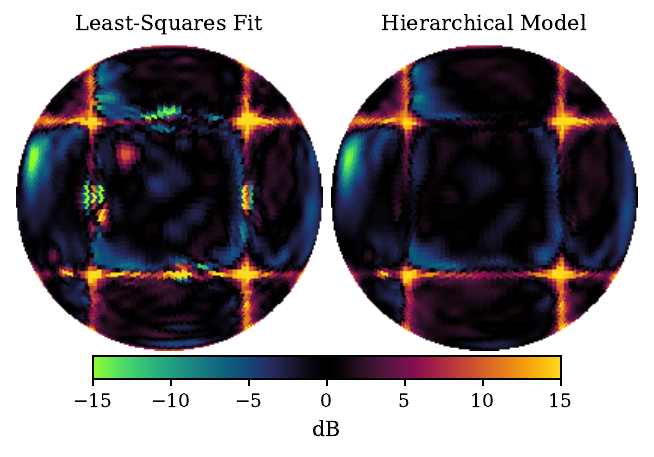}
    \caption{Comparison of least-squares fit to hierarchical model fit for third chunk of jackknife data (third row of Figure \ref{fig:chunk_fits}). }
    \label{fig:lsq_comparison}
\end{figure}

As an additional comparison to showcase the performance of the hierarchical model, we compare the best fit spline for the second chunk of data to a least-squares spline fit, equivalent to a uniform Gaussian noise model with (improper) flat priors on the spline coefficients (Figure \ref{fig:lsq_comparison}). The hierarchical model has substantially fewer spurious structures even in the main lobe where sampling is poor, and the least-squares solution has spurious structures even in the sidelobes where sampling is better. This is especially apparent when we compare these two inferred patterns to the measured ratio using the full data set in Figure \ref{fig:spline_example}. The hierarchical fit bears a much stronger resemblance to the full data set than the least-squares fit. These inferred patterns are based on the same data and basis functions, so it is clearly the statistical structure of the hierarchical model that enhances its ability to predict beams from sparse sampling patterns as compared to the more basic least-squares fit. 

\begin{figure}
    \centering
    \includegraphics[width=\linewidth]{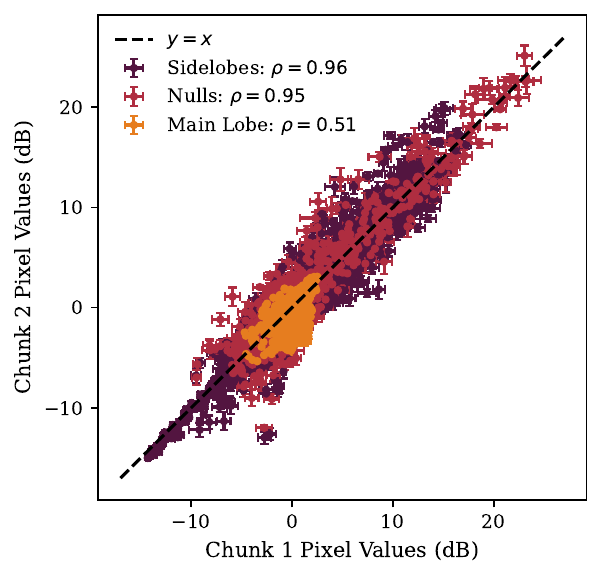}
    \caption{Pixel values for the best-fit splines from the first and second chunks of data, broken down by region and with line $y=x$ to guide the eye (i.e. the relationship where the two beams are equal). We list the region-specific correlation coefficients in the legend.}
    \label{fig:chunk_corr}
\end{figure}

The consistency of inferred beams from different chunks can be quantified by the pixel-pixel correlation coefficient. We find that the inferred beams are highly correlated in the nulls and sidelobes, but only moderately correlated in the main lobe. This is demonstrated for the first two chunks of data in Figure \ref{fig:chunk_corr}. 

Each $(x, y)$ coordinate in this plot is a mean value for a given pixel according to each chunk of data. The error bars represent uncertainties calculated using MCMC samples. That is, we mapped the spline coefficients to beam values in the HEALPix bins, and took the standard deviation of the transformed samples to obtain the uncertainties.

We broke the maps down into three regions, corresponding to our knot density choice: main lobe, sidelobes, and nulls. We then calculated weighted Pearson correlation coefficients between the pixel values for the two inferred beams, where the weights were given by the inverse of the sum of the horizontal and vertical variances. The fact that the paired pixels generally lie along the line $y=x$ speaks to the broad consistency of the inferred beam patterns from the two different chunks. They are particularly consistent in the sidelobes and nulls, where correlation coefficients exceed 0.95. The main lobe, where sampling is poor, exhibits only a correlation coefficient of 0.51. We also found these results to be consistent between different chunk pairs, with small variations in the exact correlation coefficients. 

Notice, however, that on a pixel-by-pixel basis, main pixels are several standard deviations displaced from the line $y=x$. This indicates that these are statistically inconsistent beams on a pixel-by-pixel basis. Indeed, when we generate a posterior predictive check equivalent to the right panel of Figure \ref{fig:error_stats} using the beam of one chunk and the noise map of another chunk, there are many highly significant outliers. In other words, one chunk's data is not very probable assuming another chunk's beam. Since the posterior uncertainty represents the uncertainty that may arise from noise and sampling, this means the excess variance between inferred beam patterns is unlikely to be strictly due to differences in sampling alone. Rather, we suspect that there are unchecked systematic effects in the data, about which we speculate in \S\ref{sec:discussion}. Nevertheless, the broad consistency indicated by the maps in Figure \ref{fig:chunk_fits} and quantified by the high correlations in the null and sidelobe regions demonstrate that this is an effective method for using limited data to extract deviations of the true receiving element from an idealized model. In other words, where the correlation coefficient is large, the statistical inconsistency of the inferred beams from different chunks is small compared to the size of deviations from the idealized electromagnetic simulations, and using any one of these beams in those regions will likely be a large improvement for science cases where polarization can be ignored (i.e. where a power beam is sufficient). 

\begin{figure}
    \centering
    \includegraphics[width=\linewidth]{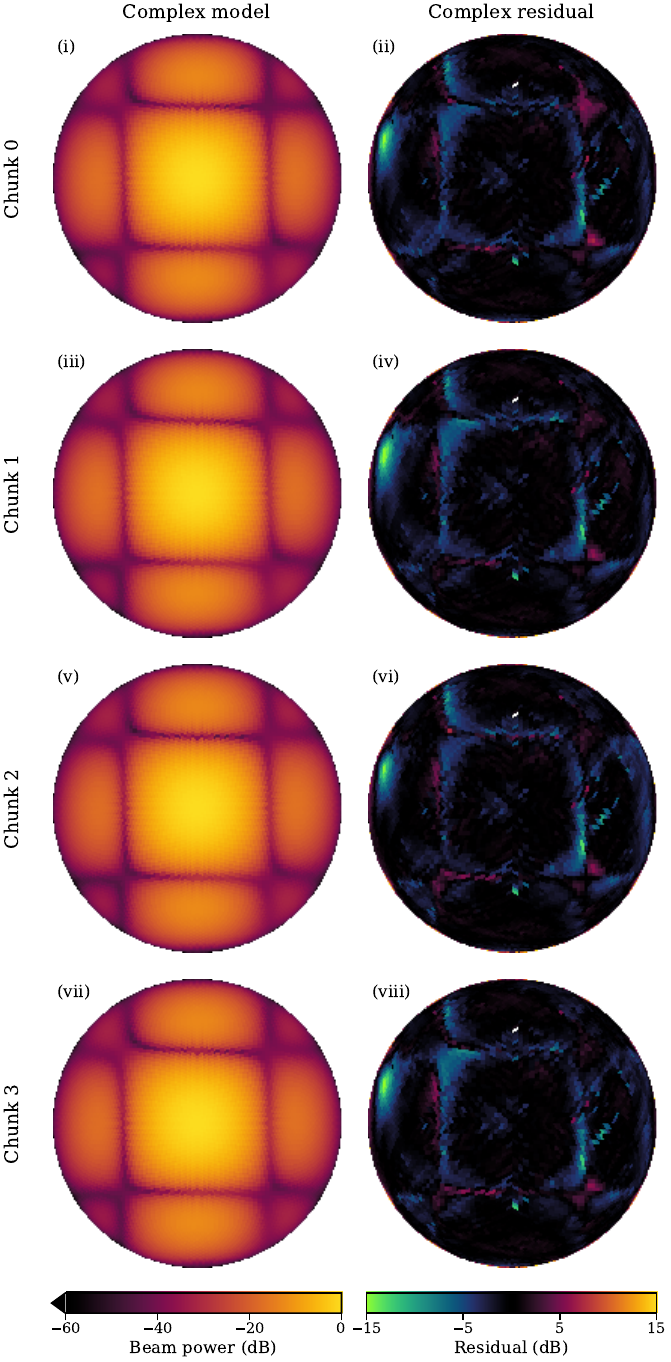}
    \caption{Temporal stability of the inferred S06XX beam. Posterior mean beams (left) and corresponding full-data residuals (right) for the complex dipole model, inferred independently from four 20-day chunks of satellite data. Residuals are defined as the chunk posterior-mean model minus the measured beam.}
    \label{fig:dipole_chunk_fits}
\end{figure}

We next apply the same temporal-stability test to the complex dipole model. Figure~\ref{fig:dipole_chunk_fits} shows the posterior-mean beam inferred independently from each 20-day chunk, together with the residual obtained when each chunk-derived beam is evaluated against the full dataset. Despite the sparse and non-identical sampling, the four inferred beams have highly consistent null and sidelobe structure. The differences are most apparent at the corners of the main lobe where three sidelobes meet. There are fainter differences in the main lobe, although they are less visually prominent than in the more flexible spline reconstructions. 

The residual maps are also largely consistent across the four fits. In particular, the western-sidelobe depression and coherent structure around the nulls remain apparent regardless of which subset is used to infer the dipole gains. The two columns of Figure~\ref{fig:dipole_chunk_fits} therefore support related but distinct conclusions: the similar inferred beams demonstrate the temporal stability of the dipole reconstruction, while the recurring residual morphology identifies reproducible structure that cannot be accommodated by direction-independent complex gains. Because each residual map is evaluated against the same full dataset, these panels do not constitute four independent detections of the residual features; rather, they show that independently inferred dipole models share the same limitations.

\begin{figure}
    \centering
    \includegraphics[width=\linewidth]{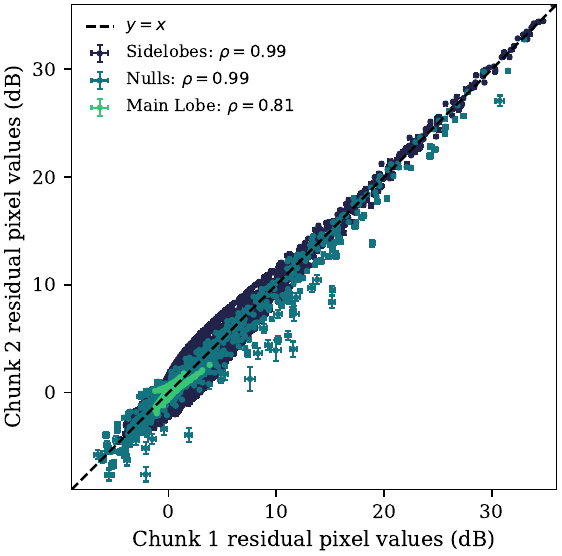}
    \caption{Pixel-wise beam residuals inferred independently from the first and second 20-day chunks of satellite data, separated by beam region. Residuals are defined as the inferred complex-gain beam minus the nominal AEE beam. The dashed line shows $y=x$, corresponding to identical inferred beam residuals, and the region-specific correlation coefficients are given in the legend.}
    \label{fig:dipole_chunk_corr}
\end{figure}

To quantify the agreement between the inferred dipole corrections without it being dominated by the common nominal beam structure, we compare their departures from the AEE model, $\Delta B_c=B_c-B_{\mathrm{AEE}}$, where $c$ denotes the 20-day chunk. Figure~\ref{fig:dipole_chunk_corr} compares these departures for the first two chunks. The inferred corrections are highly correlated in the sidelobes and nulls, with $\rho=0.99$ in both regions, while the main-lobe correlation is lower at $\rho=0.81$. These region-wide correlations demonstrate strong repeatability of the broad spatial morphology, but do not imply uniform agreement at every location. In particular, the localised differences near the corners of the main lobe, where three sidelobes meet, affect relatively few pixels and therefore do not substantially reduce the overall sidelobe and null correlations.

The lower correlation in the main lobe plausibly has two observational causes. First, the sparser and noisier sampling leaves each pixel more susceptible to differences between the satellite passes contained in each chunk. Second, the high-power main-lobe measurements are known to be affected by amplifier saturation. Although the monotonic-concave calibration introduced in \S\ref{sec:data} substantially reduces this non-linearity, it does not remove it completely. Residual saturation-related errors could therefore differ between pass ensembles and produce chunk-dependent main-lobe corrections without requiring temporal variation in the physical beam. The smaller dynamic range of the inferred corrections within the main lobe also makes its Pearson coefficient more sensitive to modest pixel-level differences than those of the null and sidelobe regions.

The higher main-lobe correlation of the dipole model compared with the spline model should therefore not be interpreted as evidence of greater accuracy. The compact dipole parametrisation couples the response across the hemisphere and permits substantially less local variation, naturally producing more similar reconstructions from different subsets. Moreover, the correlation coefficients quantify the reproducibility of the spatial morphology rather than equality of the inferred corrections. Identical chunk models would lie exactly along $y=x$ and have $\rho=1$, but a high correlation can persist in the presence of systematic offsets or differences in scale. Although the points generally follow $y=x$, some displacements exceed the posterior uncertainties. The inferred corrections are therefore broadly stable but not identical within their quoted uncertainties.

Taken together, the two jackknife analyses show that the principal departures from the nominal electromagnetic beam models can be recovered from sparse and non-identical satellite-pass ensembles. Both models recover highly reproducible null and sidelobe structure, while showing greater subset dependence in the main lobe and, for the dipole model, localised differences where several sidelobes meet. The reduced main-lobe agreement is consistent with the combined effects of sparse sampling and residual measurement non-linearity rather than necessarily indicating temporal variation of the beam. The western depression is repeatedly reconstructed by the spline model but remains in the residuals of every dipole fit, strengthening the conclusion that it represents a reproducible, direction-dependent discrepancy that cannot be described by constant per-dipole gains. Temporal repeatability does not, however, conclusively establish whether this structure originates in the receiving tile or in a persistent measurement systematic.

\section{Prospects for Future Experiments}
\label{sec:discussion}

The agreement between independent temporal subsets demonstrates that much of the measured null and sidelobe structure is reproducible under different sampling patterns. Nevertheless, the pass-shaped residuals and pixel-level differences between chunk inferences reveal systematic effects not described by the adopted noise models. The main lobe is particularly challenging: insufficient attenuation caused the RF Explorer receivers to compress its high-power response, whereas greater attenuation would have reduced sensitivity to the much weaker sidelobes and nulls. Although our receiver-specific recalibration substantially mitigates this non-linearity, the lower cross-chunk agreement suggests that residual compression-related errors may remain. Future analyses could model the measurements at the level of individual passes, accounting for quantities such as receiver configuration, received power and sub-pixel trajectory rather than absorbing all remaining variation into a per-pixel noise scale.

A more fundamental limitation is that the present measurements constrain monochromatic power beams near 137~MHz. Precision wide-band experiments of the redshifted 21-cm signal require the chromatic and polarimetric response of the instrument. Power measurements constrain quantities such as $[\mathbf{J}\mathbf{J}^{\dagger}]$ but do not uniquely determine the complex Jones matrix $\mathbf{J}(\hat{\mathbf{s}},\nu)$. The resulting loss of phase information produces the conjugate-rotation ambiguity identified in \S\ref{sssec:symmetry} and prevents a complete reconstruction of direction-dependent polarisation leakage. Application to wide-band polarimetric science therefore requires measurements across frequency and, ultimately, a phase-referenced experiment capable of constraining the full Jones response.

The growing number of satellites with detectable unintended out-of-band emissions may provide opportunistic beam measurements at several frequencies \citep{Grigg_SatRFI_2025}. Such transmitters offer repeated all-sky coverage, but their frequencies, powers and polarisation states are not under experimental control. Broadband drone measurements provide a complementary route by densely sampling the frequency dependence of the inferred dipole parameters and spline corrections under controlled conditions. These measurements could be used to construct a frequency-dependent physical model, while satellite observations provide continuous monitoring and spatial coverage at the frequencies where suitable transmissions are available. Establishing an empirical frequency evolution from a representative off-site tile would not by itself guarantee an accurate chromatic model for every installed tile, particularly where local geometry and environmental conditions matter. The frequency dependence inferred from controlled measurements must therefore be tested against satellite maps of the deployed elements.

A purpose-designed coherent satellite beacon could address several limitations of the present power-based experiment. Rather than measuring total received power, the telescope would record voltages and correlate them with a known pseudorandom signal transmitted by the beacon. Incoherent power averaging improves the detection signal-to-noise ratio only with the square root of the number of independent samples, correlation with a known waveform accumulates the signal coherently and allows the signal-to-noise ratio to grow linearly in the ideal limit. Coherent detection therefore provides an additional square-root gain over an equivalent power measurement. The beacon could therefore transmit at lower instantaneous power while retaining sensitivity to the sidelobes and nulls. This would substantially reduce the risk of saturation in the main lobe. 

Voltage measurements would also retain the phase information discarded by the present experiment. A stable phase reference would remove the specific power-beam degeneracy discussed in \S\ref{sssec:symmetry}, apart from an overall phase convention, while transmissions in known polarisation states could constrain the full Jones response. Because the pseudorandom waveform would be known in advance, it could also be identified and potentially subtracted from the recorded voltages before astronomical visibilities are formed. A coherent beacon could thereby map the chromatic, polarised beam while limiting its impact on simultaneous astronomical observations.

The two inference models developed here also suggest a natural direction for future statistical modelling. The dipole model provides a compact physical description of much of the measured deformation, but its direction-independent gains cannot reproduce all of the localised structure. Conversely, the spline model provides a more complete phenomenological reconstruction without identifying the physical mechanism responsible. A hybrid model could combine a physical AEE-based component with a regularised direction-dependent correction, allowing the data to constrain interpretable instrumental parameters while retaining sufficient flexibility to describe reproducible departures from the assumed geometry. Care would be required to maintain identifiability between the two components so that the flexible correction does not simply absorb structure that could otherwise constrain the physical model. Extending the physical component to include tile rotation, tilt, dipole displacement, non-coplanarity or ground-screen deformation may also provide more diagnostic descriptions than complex gains alone.

The dipole inference procedure is directly applicable to other MWA pointings. For each pointing, the appropriate beamformer delays can be implemented in the nominal AEE array factor and the gains inferred anew; no assumption is required that the gains inferred at zenith remain unchanged. The present spline implementation is less immediately transferable because its knot positions were selected manually from the structure of the zenith-pointed FEE beam. Other pointings or instruments would require either a sufficiently dense uniform grid or an automated prescription for placing knots where the nominal response varies most rapidly. For example, knot density could be made a function of the gradient or curvature of the nominal beam in dB, concentrating flexibility around nulls and other sharp structure while retaining a sparse representation in smoother regions. Such a construction would also allow for adaptive knot construction as a function of frequency. Selecting knots from the nominal electromagnetic model, rather than from the measurements themselves, would prevent pass-dependent artefacts from influencing the model architecture and simplify statistical inference. Developing and validating such an adaptive construction remains an important direction for future work. However, we also remark that much of the difficulty with knot placement arises from the constraint of positive-definiteness when working with the power beam (i.e. from working in logarithmic units), and that working with full Jones measurements could in principle be done in linear units where the beam smoothness is more homogenous. 

Together, these developments would extend the present demonstration from monochromatic power-beam reconstruction towards repeatable, chromatic and polarimetric characterisation of deployed receiving elements. Opportunistic satellites can provide persistent all-sky sampling, controlled drone measurements can constrain frequency evolution, and coherent beacons can supply the phase and polarisation information required for Jones-matrix inference. Combined with physical and spatially flexible statistical models, these measurements offer a path towards beam models that can be propagated directly into the calibration and foreground-control requirements of future high-dynamic-range radio experiments.

\section{Conclusions}
\label{sec:conclusion}

We have developed two complementary Bayesian approaches for inferring continuous MWA beam patterns from irregular and incomplete satellite sampling. We first returned to the raw measurements of \citet{Chokshi2021}, avoiding the correlations introduced by their previous interpolation and smoothing, and derived receiver-specific corrections for the non-linear response of the RF Explorer spectrum analysers. The reprocessed measurements were then binned onto a HEALPix grid with per-pixel uncertainties that account for both within-pass noise and variations between contributing satellite passes.

The S06XX measurements exhibit coherent departures from both the nominal AEE and FEE predictions, most prominently through shallower and displaced nulls, a broad depression in the western sidelobe, and large-scale gradients across the sidelobes. The dipole model represents much of this deformation using a compact set of per-dipole gains. Allowing the amplitudes to vary reduces the residual RMS from 7.889 to 2.945~dB, while the statistically preferred complex model further reduces it to 2.890~dB and lowers the MAD from 1.188 to 1.015~dB. The power-only measurements leave two symmetry-related complex-gain configurations that are physically distinct in parameter space but predict effectively identical power beams. The inferred gains should therefore be interpreted as effective descriptions of the measured deformation rather than unique diagnoses of electronic gain errors.

The hierarchical spline model provides a complementary reconstruction relative to the FEE beam. It recovers additional direction-dependent structure around the nulls and throughout the sidelobes while generally leaving isolated pass-shaped artefacts in the residuals. In particular, the spline repeatedly reconstructs the broad western depression that remains in the residuals of the dipole model, demonstrating that this feature requires spatial flexibility beyond direction-independent complex gains. Agreement between the two models identifies structure that is supported under substantially different parameterisations, while their differences distinguish what can and cannot be represented within the nominal AEE geometry.

A central result of this work is that a continuous full-hemisphere beam pattern can be reconstructed from only 20 days of satellite observations. Each of four independent 20-day subsets contains approximately $20\%$ of the full dataset and provides a distinct, incomplete sampling pattern. Nevertheless, every subset recovers the principal null and sidelobe morphology. Correlations between the first two spline reconstructions reach 0.95 and 0.96 in the nulls and sidelobes, respectively, while the corresponding dipole corrections have correlations of 0.99 in both regions. Agreement is weaker in the main lobe, where the measurements are sparser and more susceptible to residual receiver non-linearity. Similar behaviour for S10YY (Appendix \ref{sec:ant_10_demo}) demonstrates that the reconstruction is not unique to a single tile or polarisation.

The present results remain limited to monochromatic power beams near 137~MHz. Temporal repeatability also cannot by itself establish whether a recovered feature originates in the receiving tile or in a persistent measurement systematic. Extending this approach to chromatic, phase-referenced voltage measurements would enable inference of the full Jones response required by wide-band polarimetric experiments. Nevertheless, the recovery of consistent full-hemisphere beam patterns from independent 20-day observing intervals demonstrates that sparse satellite passes can provide the information required for practical and repeatable characterisation of deployed receiving elements.

\section*{Acknowledgements}

We are grateful to Robert Pascua, Bryna Hazelton, Miguel Morales and Jonathan Pober for insightful discussions. This scientific work makes use of the Inyarrimanha Ilgari Bundara Murchison Radio-Astronomy Observatory, operated by CSIRO. We acknowledge the Wajarri Yamatji people as the traditional owners of the Observatory site.

Large language models (Claude Sonnet 5 and GPT 5.6 Sol) were used to assist with writing, debugging and refactoring portions of the analysis code, and to improve clarity and phrasing in parts of the manuscript. The statistical methods and validation checks were developed independently of these tools. Every reported number can be regenerated from scripts and notebooks committed to the repository, and the authors take full responsibility for the content of this paper.

\section*{Data Availability}

All pre-processed pass data, including the recalibrated passes, are available from \url{https://github.com/amanchokshi/embers-passes}.

\appendix

\section{Demonstration with another tile}
\label{sec:ant_10_demo}

Here we repeat a subset of the inference and diagnostic plots for a second tile--polarisation combination, S10YY. This provides a direct demonstration that the methodology and the broad inferred morphology are not unique to tile S06XX, although we have not repeated the complete analysis for every measured tile. Visual inspection of the satellite beam maps and their residuals for all 14 tiles in both polarisations nevertheless reveals several recurring qualitative features, including shallower nulls, square-like residual structure around the main lobe, and broad gradients across the sidelobes. These similarities suggest that such departures are not exclusive to the two examples analysed here, but establishing their prevalence quantitatively would require applying the full inference to the wider tile sample.

\begin{figure*}
    \centering
    \includegraphics[width=\linewidth]{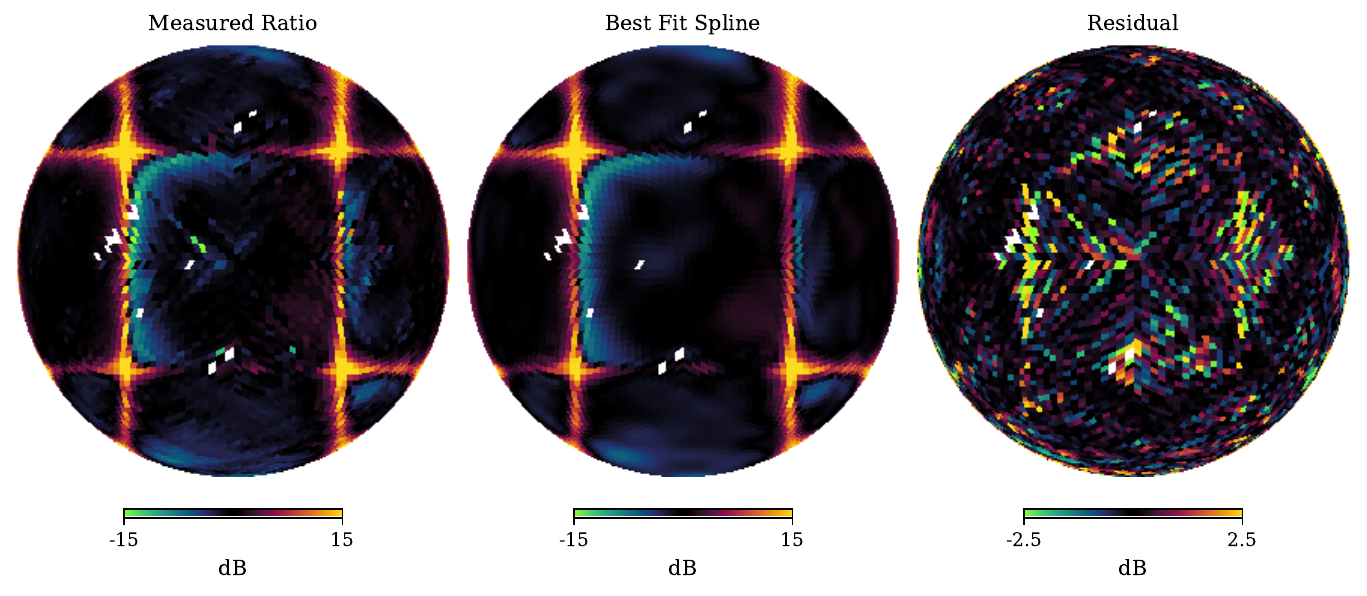}
    \caption{Equivalent to Figure \ref{fig:spline_example} but with Tile S10YY.}
    \label{fig:tile_10_spline_example}
\end{figure*}

Figure \ref{fig:tile_10_spline_example} shows the best fit spline model using the full data set along with the residuals. Notably, the shallow nulls and null displacement features also present for this tile and polarization, and the main lobe residuals show pass-shaped artefacts in them. The FEE beam for the XX vs. YY polarization are rotated by 90 degrees azimuthally from one another. Examining the Northern and Southern sidelobes of tile S10YY, we do see coherent depressions of the beam response towards the horizon similar to tile S06XX, however they are not as pronounced and therefore more difficult to determine whether they are spurious. Interestingly in the residuals, it is now the vertically running null lines that demonstrate the oscillatory residuals. This and the 90 degree rotation symmetry of the FEE beam make this feature seem like a genuine property of the tile beams. As discussed in the main text, this may be a feature of tile-tile coupling in the interferometer.

\begin{figure*}
    \centering
    \includegraphics[width=\linewidth]{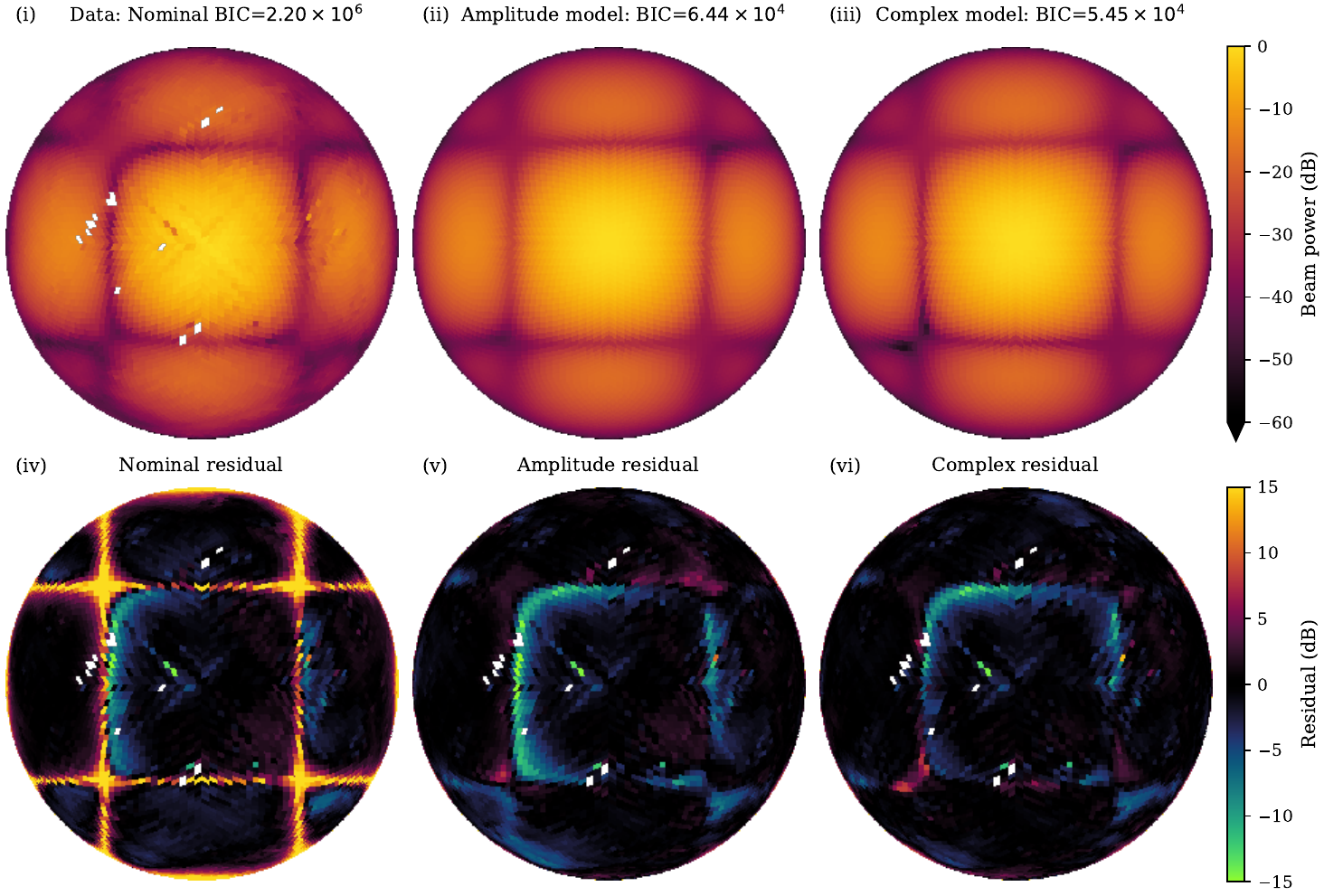}
    \caption{Equivalent to Figure \ref{fig:dipole_beam_comparison} but with Tile S10YY.}
    \label{fig:tile_10_dipole_example}
\end{figure*}

Figure~\ref{fig:tile_10_dipole_example} shows the corresponding full-data comparison for the dipole model. Allowing the dipole amplitudes to vary reduces the BIC from $2.20\times10^6$ for the nominal AEE to $6.44\times10^4$, while allowing complex gains further reduces it to $5.45\times10^4$. The phase degrees of freedom are therefore strongly supported after accounting for the additional model complexity, reproducing the model ordering found for S06XX. Both inferred dipole models capture much of the broad deformation of the nominal beam, including the shallower and displaced null structure, with the complex model providing the better description. Nevertheless, coherent residuals remain around the nulls and in regions of the sidelobes. The S10YY results therefore reinforce the conclusion that direction-independent per-dipole gains provide a compact representation of much of the measured deformation but cannot reproduce all of its localised structure.

\begin{figure}
    \centering
    \includegraphics[width=\linewidth]{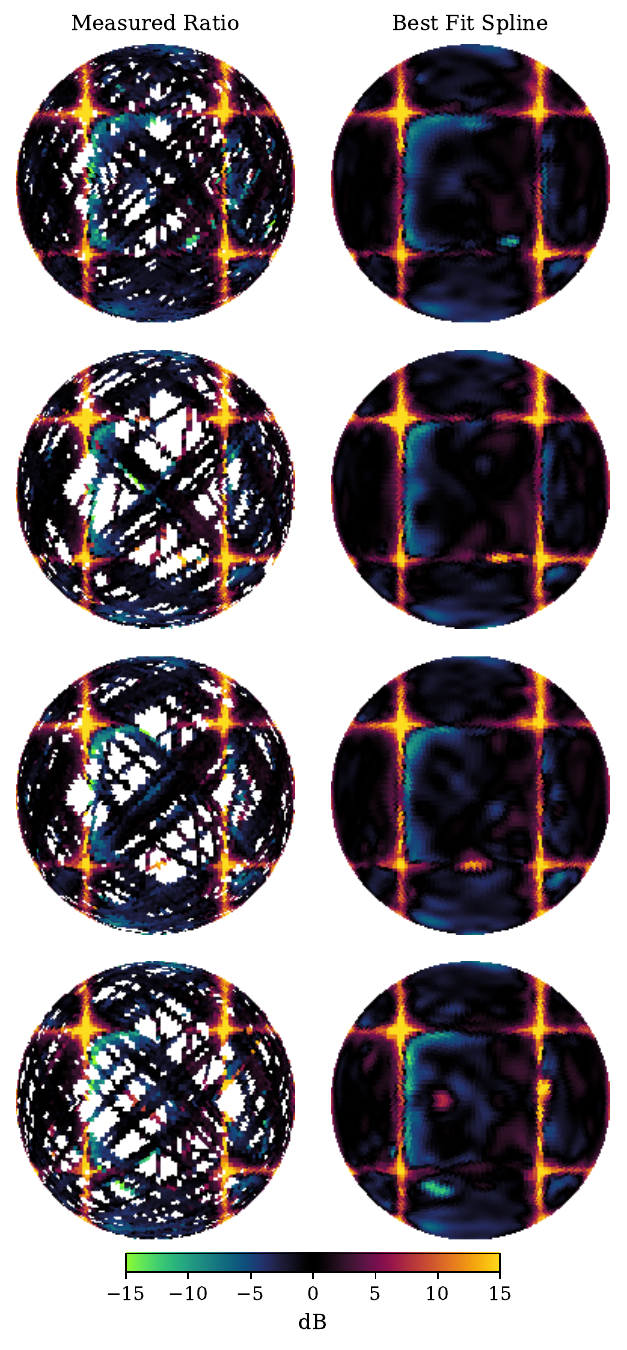}
    \caption{Equivalent to Figure \ref{fig:chunk_fits}, but with tile S10YY.}
    \label{fig:tile_10_chunk_fits}
\end{figure}

\begin{figure}
    \centering
    \includegraphics[width=\linewidth]{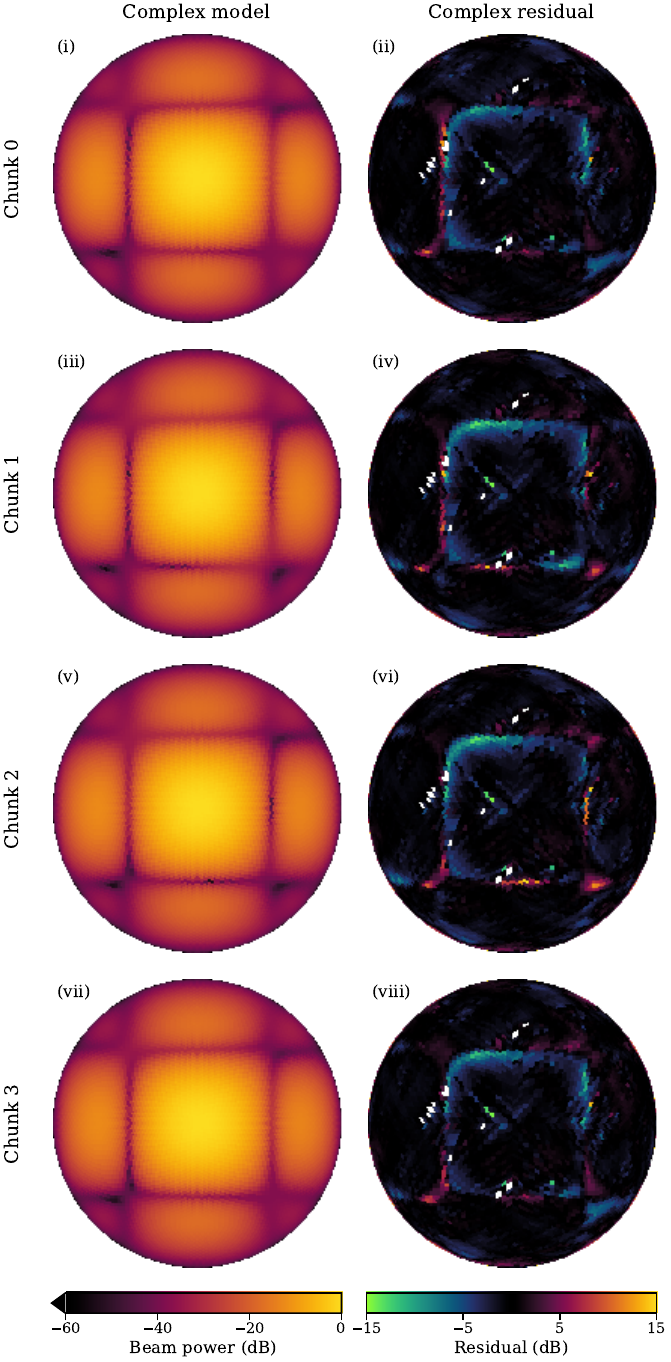}
    \caption{Equivalent to Figure \ref{fig:dipole_chunk_fits}, but with tile S10YY.}
    \label{fig:tile_10_dipole_chunk_fits}
\end{figure}

Figures~\ref{fig:tile_10_chunk_fits} and \ref{fig:tile_10_dipole_chunk_fits} repeat the temporal-subset test for the spline and complex dipole models, respectively. The S10YY measurements are more sparsely sampled than those of S06XX, and the coverage differs substantially between the four 20-day chunks. Nevertheless, the spline reconstructions recover broadly consistent null and sidelobe morphology, including the northern and southern depressions, although the southern null is less consistent and the western null is rather sparsely sampled in some chunks. The independently inferred dipole beams are likewise visually stable across the four subsets. When each chunk-derived dipole beam is evaluated against the full dataset, coherent residual structure recurs around the nulls and in the sidelobes. As in the S06XX analysis, the similar inferred beams demonstrate the stability of the dipole reconstruction, whereas the recurring residual morphology identifies stable limitations of the direction-independent gain model. Because these residual panels use the same full dataset, they should not be interpreted as independent detections of the residual features.

Together, the S10YY results reproduce the principal qualitative conclusions obtained for S06XX under a different tile, polarisation, and sampling pattern. Both models recover broad, repeatable departures from the nominal electromagnetic predictions; the dipole model captures much of this structure through a compact physical parametrisation, while the spline accommodates additional direction-dependent morphology. The recurrence of these behaviours motivates applying the complete analysis to the wider tile sample.

\section{Bayesian Information Criterion}
\label{sec:BIC}

Here we discuss the Bayesian Information Criterion (BIC) and show its use in comparing different spline models. The BIC is a model selection criterion introduced by \citet{BIC}. It is given by
\begin{equation}
    \mathrm{BIC} = k\ln n - 2\ln\hat{L},
    \label{eq:BIC}
\end{equation}
where $k$ is the dimensionality of the model, $n$ is the dimensionality of the data, and $\hat{L}$ is the maximum likelihood value. With this sign convention, it approximates the negative marginal likelihood, and lower scores imply more supported models. The marginal likelihood is a robust metric for model selection that naturally penalizes extra parameters unless they provide sufficient improvements in fit \citep{Mackay2003}. This is manifest in Equation \ref{eq:BIC}, where there is a linear dependence on parameter dimension that directly competes with the log-likelihood. 

We note that in our calculations for the spline model, we have actually evaluated the likelihood at the MAP solution since this was substantially more convenient to compute. It is therefore approximate. However, we expect that this produces a negligible difference since this was calculated with the full data set, meaning each spline basis function is well-sampled by the data and the hierarchical structure of the model does little to influence the parameter estimate. For the sparser data sets (e.g. the 20-day chunks), some spline coefficients may be poorly constrained by the data and therefore relying on the hierarchical structure for a fit, which could produce a significant discrepancy between the MAP estimate and maximum likelihood estimate of the parameters.

\begin{figure}
    \centering
    \includegraphics[width=\linewidth]{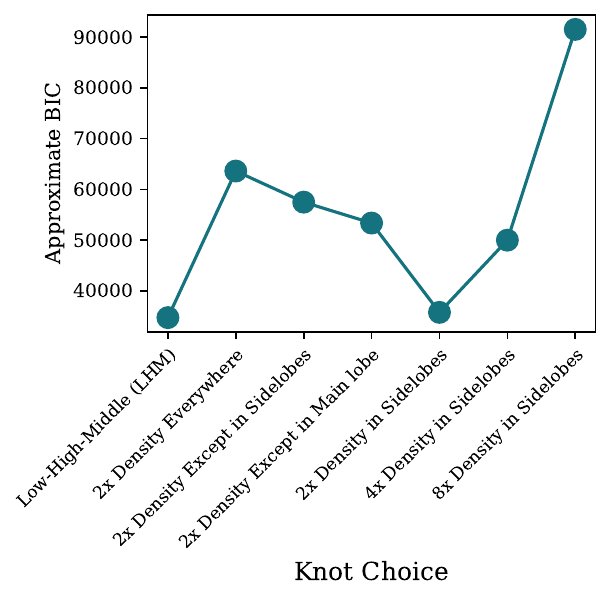}
    \caption{Evaluated BIC for different choices of spline model. Densities are expressed with respect to the leftmost label, which is itself expressed in the region order (main lobe, nulls, sidelobes). We chose the LHM model for all spline-based inference in this work.}
    \label{fig:spline_BIC}
\end{figure}

We applied the approximate BIC to a sequence of knot choices. Figure \ref{fig:spline_BIC} shows the results. The left-most choice, designated ``Low-High-Middle'' (LHM), is the choice we used for all demonstrations in this work. The naming convention refers to the relative density of knots in the main lobe, null, and sidelobe regions. All other model labels are named with respect to the densities of the knots in this model. For example, ``2x Density Everywhere'' means that the grid defining the knots has twice as many points everywhere, whereas ``2x Density in Sidelobes'' means only the grid of knots in the sidelobe regions received twice as many points. 

Generally we found that increasing the knot density beyond the LHM model resulted in worse performance, except when we increased knot density in the sidelobes. In this alternate model, fit at the horizon is improved, resulting in a decrease in the number of outliers as observed in Figure \ref{fig:error_stats}. However, since it scored essentially the same on the BIC, we decided that this extra parametrization was unnecessary. 

\section{Demonstration of hierarchical structure in spline coefficients}
\label{sec:hierarchical_structure}

\begin{figure}
    \centering
    \includegraphics[width=\linewidth]{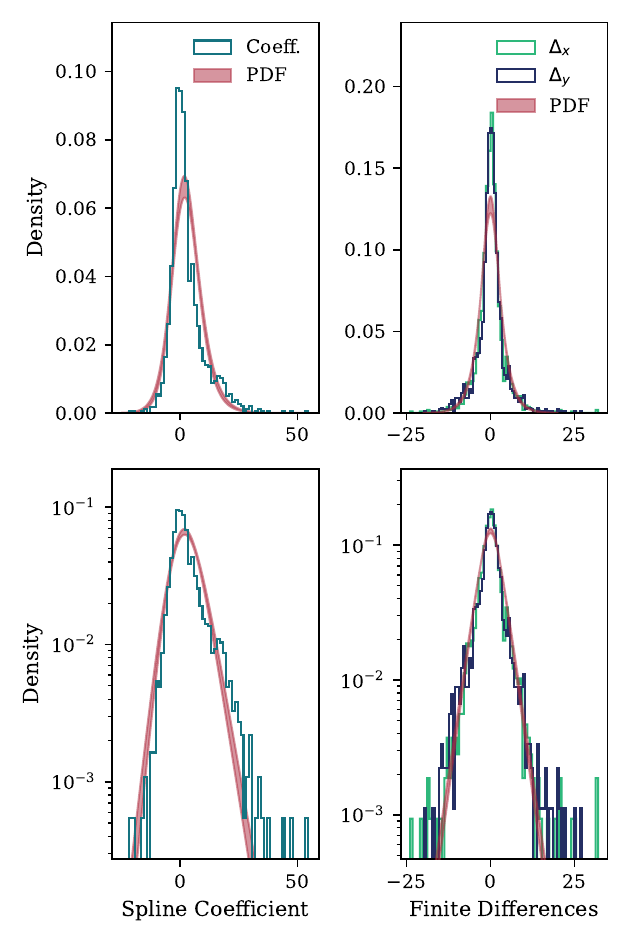}
    \caption{Histograms comparing best-fit spline coefficients for tile S06XX (left) and finite differences (right) on linear (top) and logarithmic (bottom) y-axes to their inferred probability density functions (PDFs) from the hierarchical model. The fill for the PDFs shows 5 posterior standard deviations in the value of the PDF based on uncertainty in the shape and scale parameters.}
    \label{fig:hierarchical_hist}
\end{figure}

We demonstrate that the spline coefficients and their finite differences roughly follow the distributions used in the hierarchical model. We show this in Figure \ref{fig:hierarchical_hist}, where the best-fit (posterior mean) spline coefficients and their finite differences for tile S06XX are shown alongside their inferred probability density functions (PDFs) based on the skew logistic and soft Laplace distributions. The scale and shape parameters of these distributions are extremely well-constrained; uncertainties in the value of the PDFs are enhanced by a factor of 5 for visual clarity. To calculate these uncertainties, we evaluate the PDF for each posterior sample, take the standard deviation of the evaluated PDFs at each point, and display a region corresponding to the mean plus or minus five standard deviations.

In both cases of the coefficients and the finite differences, the observed distributions are somewhat more peaked than the inferred distributions. However, the overall shape of the tails is well-matched for the finite differences. The fit for the spline coefficients is rougher, but the negative tail fits better than the positive one. While we believe these are sufficiently specified distributions to avoid complications from having overly thin tails (e.g. compared to a Gaussian alternative), future work could focus on tailoring the distributions for more accurate regularization of the spline coefficients. This particular choice may also only be appropriate for the zenith pointing of MWA beams, and we have no way of knowing whether beams of other instruments will obey these distributions even approximately. However, a future analyst can essentially decide on this iteratively by observing data, fitting it to the model, and refining from there with a new choice of hierarchical distribution.

\bibliographystyle{paslike}
\bibliography{bibliography}

\end{document}